\documentclass{statsoc}
\usepackage{babel}
\usepackage{graphicx}
\usepackage[a4paper,margin=2cm]{geometry}
\def\mm#1{\ensuremath{\boldsymbol{#1}}}
\usepackage{amsmath,amssymb}
\usepackage{graphicx}
\usepackage[colorlinks=true, allcolors=blue]{hyperref}

\usepackage{array}
\newcolumntype{L}[1]{>{\raggedright\let\newline\\\arraybackslash\hspace{0pt}}p{#1}}

\title[UK Health Index]{Assessments and developments in constructing a National Health Index for policy making, in the United Kingdom}
\author[Freni-Sterrantino {\it et al.}]{ Anna Freni-Sterrantino}
\address{The Alan Turing Institute, London, UK}
\email{afrenisterrantino@turing.ac.uk}

\author{Thomas P. Prescott}
\address{The Alan Turing Institute, London, UK}

\author{Greg Ceely}

\address{Office for National Statistics}

\author{Myer Glickman}

\address{Office for National Statistics}

\author{Chris Holmes}
\address{The Alan Turing Institute, London, UK}
\email{cholmes@turing.ac.uk}

\begin{document}

\begin{abstract}
 

Composite indicators are a useful tool to summarize, measure and compare changes among different communities. The UK Office for National Statistics has created an annual England Health Index (starting from 2015) comprised of three main health domains - lives, places and people - to monitor health measures, over time and across different geographical areas (149 Upper Tier Level Authorities, 9 regions and an overall national index) and to evaluate the health of the nation. The composite indicator is defined as a weighted average (linear combination) of indicators within subdomains, subdomains within domains, and domains within the overall index. 
The Health Index was designed to be comparable over time, geographically harmonized and to serve as a tool for policy implementation and assessment. 
 
We evaluated the steps taken in the construction, reviewing the conceptual coherence and statistical requirements on Health Index data for 2015-2018. To assess these, we have focused on three main steps: correlation analysis at different index levels; comparison of the implemented weights derived from factor analysis with two alternative weights from principal components analysis and optimized system weights; a  sensitivity and uncertainty analysis  to assess to what extent rankings depend on the selected set of methodological choices. Based on the results, we have highlighted features that have  improved statistical requirements of the forthcoming UK Health Index.

\end{abstract}

\section{Introduction}
\label{sec:intro} 

A composite index (CI) is a way to summarize several indicators in one number and provide a tool for policy-making.  Besides the known health-related indices like Healthy  Life Expectancy \cite{van1996policy} or Disability-Adjusted Life Years \cite{hyder2012measuring,soerjomataram2012estimating}, in the United Kingdom (UK) there has been a long tradition of health-related indices; the first `Health Index' was developed in 1943 as a surveillance system for population health at national level, based on mortality and morbidity annual data \cite{sullivan1966conceptual}. Kaltenthaler et al. \cite{kaltenthaler2004population}, in their systematic review conducted in  2014,  evaluated  17  population level health indexes  and found that three  were composed for the UK population. The `Health and material deprivation in Plymouth' \cite{abbott1992health} a modification of Townsend's `Overall Health Index' \cite{townsend1988indicator} and the most popular `Index of Multiple Deprivation' \cite{department2000indices}. However none of them or any of the other health-population indexes  seemed to fulfil the desiderata for a health index:  proper health coverage indicators; routinely collected and updated data; indices at local and national  level; and  statistical coherence. These findings were later confirmed  by Ashraf et al.~\cite{ashraf2019population} in a systematic review. They concluded that 
 most of the indices  measured  population’s overall health outcomes, but only few gave focus to specific health topics or the health of specific sub-populations.  They urged the development of population health indices that can be constructed systematically and rigorously, with robust processes and sound methodology. 
 
Recently, to fill this gap, the Office for National Statistics of the UK (ONS)  developed an annual (experimental)  composite index to quantify health in England, to track changes in health across the country and to compare health measures across different population subgroups. 

The Health Index (HI) expands the  WHO definition of health: `a state of complete physical, mental and social well-being and not merely the absence of disease and infirmity' \cite{grad2002preamble}, to include  health determinants that are known to influence people's health.
Therefore, the HI is characterized by three main domains:
Healthy \emph{People},
Healthy \emph{Lives}
and Healthy \emph{Places},
split across 17 subdomains, for a total of 58 indicators.
For example, life expectancy and the standardized number of avoidable deaths define the subdomain `Mortality' and prevalence at Upper Tier Local Authority (UTLA) level of dementia, musculoskeletal, respiratory, cardiovascular, cancer and kidney conditions define the subdomain `Physical health conditions' within the Healthy People domain.
Healthy Places is structured over 14 indicators (access to public and private green space, air and noise pollution, road safety, etc.) split in 5 subdomains: Access to green space, Local environment, Access to housing, Access to services and Crime. 

The construction of a new composite indicator is a lengthy process that takes into account several steps and choices. From the wide literature on composite indicators~\cite{barclay2019problem,freudenberg2003composite,jacobs2004measuring}, it emerges that there is no gold-standard, with every method having its own drawbacks and advantages ~\cite{greco2019methodological} relative to the purpose of each CI and its future use in policy making. 

In recent years,  extensive work  was carried out by many institutions, such as
Eurostat~\cite{Euro_Part1Term},
the Organisation for Economic Co-operation and Development (OECD)~\cite{joint2008handbook},
the Joint Research Centre (JRC)~\cite{saisana2002state}
and specific working groups at the European Commission~\cite{COIN}, to provide statistical guidance on CI construction.
The cumulative effort has provided a framework to define CI principles~\cite{nardo2005tools}, outlining the essential steps, introducing sensitivity and uncertainty analysis as a core part of composite indicators~\cite{saisana2005uncertainty} and  advancing composite indicators methodology~\cite{munda2005constructing}.

With no current unanimous approved checklist for evaluating  composite indicators, we relied on two main sources to guide us into assessing the Health Index. 
The first is based on the COIN step-list from the JRC~\cite{COIN}, which includes observations from the OECD handbook~\cite{joint2008handbook}. These elements provide a framework that will guide us on the statistical (quantitative) methodological choices and statistical analysis. The second  source is based on previous work carried out in an audit format by the JRC composite indicators expert group \cite{saisana2012sustainable, caperna2022jrc}, where they have evaluated other composite indicators. 

In this paper, in an effort to fulfill transparency requirements, we evaluated the steps taken and arising issues that come into the design of the ONS HI. We highlight areas of improvement or which warrant further investigation,  based on our findings, aiming for a statistically and conceptually coherent index, that will be integrated in the future HI release. 
This paper is structured as follows. We start by describing the beta ONS HI  for 2015-2018 structure and steps taken in its construction, in section \ref{sec:example}. In section \ref{sec:corr}, we provide an in-depth correlation analysis which will be useful for the weights system selection that we introduce in section \ref{sec:weights}.  The index validity is evaluated by sensitivity and uncertainty analysis in section \ref{sec:sens}.  At the end of each section we conclude with features that could be improved or are worthy of further considerations. Finally, we provide discussion and conclusions, in section \ref{sec:conc}.

\section{The ONS Health Index}
\label{sec:example}

\begin{figure}
\centering
\includegraphics[width=0.95\textwidth]{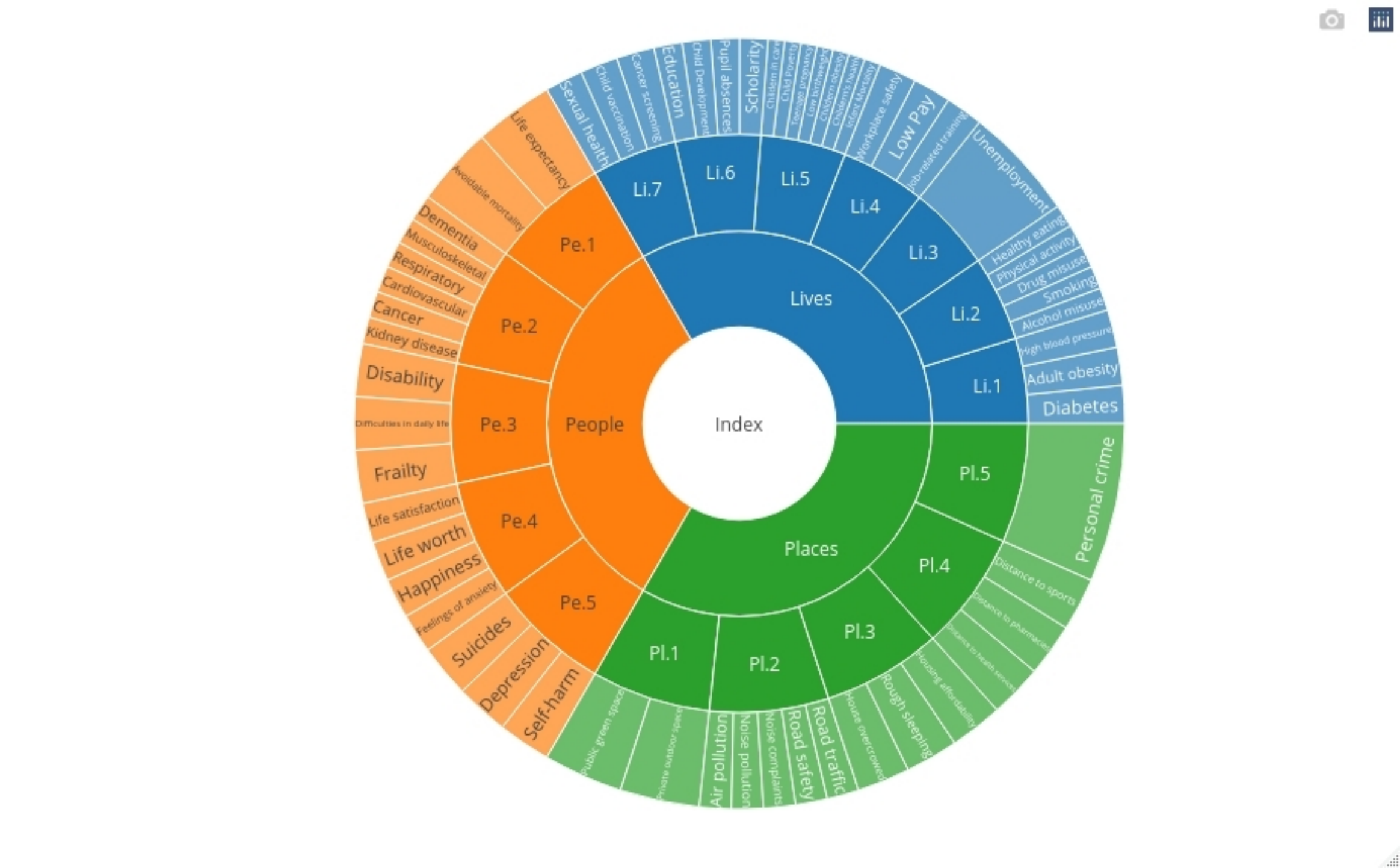}
\caption{\label{fig:newplot}The Health Index structure.}
\end{figure}

The ONS Health Index (HI) is a composite index (CI) structured in   three main domains: `Healthy People', `Healthy Lives' and `Healthy Places', see Figure  \ref{fig:newplot}. 
These domains are based on 17 subdomains, which are in turn based on 58 indicators, collected for the 149 Upper Tier Level Authorities (UTLA) in England, from 2015 to 2018. See Table~\ref{tab:tab1} for full indicator and subdomain detailed descriptions (see  also Table 1 in Supplementary Material). 
The choice of the indicators, and the definition of the 17 subdomains and three domains, were based  on a comprehensive review of contents of existing indices and frameworks; cross-referenced with existing accepted definitions of health; and then consulted on by an expert group with members from central government, local organisations, think tanks and academia to evaluate the proposal\cite{ONS}. The methodology was based on the 10 steps reported in the COIN guidance promoted by the European Joint Research Center~\cite{COIN}.
After collating raw data for the indicators at UTLA level, the steps taken to construct the Health Index were:
\begin{enumerate}
\item data imputation;
\item data treatment and normalization;
\item subdomain weights computation for factor analysis;
\item arithmetic aggregation with equal weights across subdomains and domains.
\end{enumerate}
The index is computed for each UTLA, aggregated geographically to correspond to English regions, and further aggregated into an overall national figure.
The index values are calculated for each year from 2015 to 2018 inclusive, with a normalised value anchored at the baseline year 2015.
Full details are provided in Supplementary Material (SM).

\begin{table}
\caption{\label{tab:tab1}Health Index structure: domains, subdomains and indicators}
\centering
\fbox{%
\begin{tabular}{L{0.28\textwidth}L{0.28\textwidth}L{0.28\textwidth}}
\multicolumn{3}{c}{Health Domains:} \\
People (Pe) &
  Lives (Li) &
  Places (Pl) \\
  \hline
\textbf{Pe.1} Mortality:  life expectancy, avoidable deaths &
  \textbf{Li.1} Physiological risk factors: diabetes, overweight and obesity in adults, hypertension &
  \textbf{Pl.1} Access to green space: public green space, private outdoor space \\
\textbf{Pe.2} Physical health conditions: dementia, musculoskeletal conditions, respiratory conditions, cardiovascular conditions, cancer, kidney disease &
  \textbf{Li.2} Behavioural risk factors: alcohol misuse, drug misuse, smoking, physical activity, healthy eating &
  \textbf{Pl.2} Local environment: air pollution, transport noise, neighbourhood noise, road safety, road traffic volume \\
\textbf{Pe.3} Difficulties in daily life: disability that impacts daily activities, difficulty completing activities of daily living (ADLs), frailty &
   \textbf{Li.3} Unemployment: unemployment &
  \textbf{Pl.3} Access to housing: household overcrowding, rough sleeping, housing affordability \\
\textbf{Pe.4} Personal well-being: life satisfaction, life worthwhileness, happiness, anxiety &
   \textbf{Li.4} Working conditions: job-related training, low pay, workplace safety &
  \textbf{Pl.4} Access to services: distance to GP services, distance to pharmacies, distance to sports or leisure facilities \\
\textbf{Pe.5} Mental health: suicides, depression, self-harm &
   \textbf{Li.5} Risk factors for children: infant mortality, children’s social, emotional and mental health, overweight and obesity in children, low birth weight, teenage pregnancy, child poverty, children in state care &
  \textbf{Pl.5} Crime: personal crime \\
 &
   \textbf{Li.6} Children and young people’s education: young people’s education, employment and training, pupil absence, early years development, General Certificate of Secondary Education achievement &
   \\
 &
 \textbf{Li.7} Protective measures: cancer screening, vaccination coverage, sexual health &  \\
\end{tabular}}
\end{table}

The Health Index is built starting from a tensor $\mathcal X$ of raw data, with elements $x_{cit}$. Here, each $c \in C$ is an upper tier local authority (UTLA), for the set $C$ of $|C|=149$ UTLAs;
each $i \in I$ is an indicator, for the set $I$ of $|I|=58$ indicators;
and each $t \in T = \{2015, 2016, 2017, 2018\}$ denotes the year.
We are also given a partition of the set of UTLAs, $C$, into a set $R$ of $|R|=9$ regions, $r \in R$, which are disjoint subsets $r \subseteq C$ of UTLAs.

\subsection{Data Imputation} 
We first note that $\mathcal X$ is missing data, which needs to be imputed.
Missing data was of two types: either an indicator value for a given year is completely missing for all UTLAs (see Table 2 in SM), or missing only in a subset of UTLAs.  
Briefly, if an indicator value for only one year was available, such as for `access to green space', the values were imputed to be constant across all four years.
If an indicator value is missing for a given year but available before/after, then the value was the average of the years either side of the missing year.
If an indicator value is missing and only the year before or after was available then the value would be imputed with that of the closest year. 
Full details of the data imputation are provided in the supplementary material. 

\subsection{Data treatment and normalization}

Once the missing data has been imputed, the completed tensor $\mathcal X = (x_{cit})$ is decomposed into $|I|=58$ flattened data sets, $\mm{X}_i = \{ x_{cit} : c \in C,t \in T \}$ for each $i \in I$.
Using the data transformations $f_i$ listed in Supplementary Table 3 for each indicator, $i$, the raw indicator data is transformed to $\mm Y_i = \{ y_{cit} = f_i(x_{cit}) : c \in C,t \in T \}$.
The assignment of each transformation, $f_i$, to an indicator, $i$, is selected to minimise the absolute values of skewness and kurtosis of $\mm Y_i$, aiming for absolute skewness $\leq 2$ and absolute kurtosis $\leq 3.5$. 
By minimising (absolute) skewness and kurtosis, we aim to ensure that the transformed data $\mm Y_i$ is approximately normally distributed.
For 18 indicators, the skewness and kurtosis of $\mm X_i$ were optimal, 40 indicators have been transformed and  of these 18  have been log-transformed (see Table  3 in SM).

The normalization step in the ONS Health Index accounts for time and geography, and allows indicators to be compared on the same scale, weighting by the UTLA populations.
The normalization transforms elements $y_{cit}$ of $\mathcal Y$ into z-scores,
\[
z_{cit} = (-1)^{\delta_i} \left[ \frac{y_{cit} - \mu_i}{\sigma_i} \right],
\]
which then define the elements of the tensor $\mathcal Z = (z_{cit})$.
For each indicator, $i$, we specify $\delta_i = 0$ or $\delta_i = 1$ to ensure that larger positive values for $z_{cit}$ correspond to improved health, a property which we term as being \emph{health directed}. 
Note that the mean and standard deviation $\mu_i$ and $\sigma_i$ for each indicator, $i$, are taken to be the population-weighted mean and standard deviation of $y_{cit}$ for the chosen baseline year across UTLAs $c \in C$, fixing $t=2015$. 
Finally, given the $z$-scores $z_{cit}$ forming the tensor $\mathcal Z$, the ONS Health Index presents the $z$-scores as Health Index values, 
\[
h_{cit} = H(z_{cit}) = 100 + 10 z_{cit},
\]
which are translated and rescaled $z$-scores, such that $h_{cit} = 100$ means that the transformed value, $y_{cit}$, for indicator $i$ in the UTLA $c$ in year $t$ is equal to the weighted mean, $\mu_i$. 

\subsection{Subdomain weights computation: a time-series factor analysis}

The ONS has chosen to compute weights using a time-series factor analysis. The fundamental assumption of factor analysis is that there is a latent factor that underpins the variables in a group. This translates to this level of
the Health Index: ONS assumed that there is a single unobserved variable that
underpins the indicators within each subdomain.
Highly correlated indicators within each
subdomain could lead to double counting in the index, so factor analysis directly addresses this issue, accounting for
the correlation between indicators in their implied weights \cite{decancq2013weights}.

To maintain the same weights for all the years considered (2015-18) a time-series factor analysis was applied. The rationale was to ensure that, by accounting for all the years jointly, they would change with each additional
year of data. As such, the weights would need to be calculated for a set time
period, e.g. 2015 to 2019, and these weights would be held constant until a
review date. This assured that  (i) the indicators selected matched the underlying factor (subdomains) over time; (ii) and then the factor loadings were scaled and used as data-driven weights.  

In practice,  from the normalized data $\mathcal Z_{CT} = (z_{ct})$ are collapsed by year and then rescaled to (0,1), next 
given $d \in D$,  a factor analysis on the indicators $i \in d$ was carried out and the weights were chosen as the first loading factor, taken in absolute value.
The weights $w_i$ for indicators $i \in I$ are chosen by running  factor analysis for each subdomain, $d \in D$, in turn, allowing for one factor estimated using a maximum likelihood method. 
For example, for a subdomain $d = \{ i_1, i_2 \}$ comprised of two indicators, suppose the factor loadings are 0.5 and 0.75.
We would then set the weights $w_{i_1} = 0.4$ and $w_{i_2} = 0.6$. In supplementary material, we address the weights constraints taking into account the different aggregation levels.

\subsection{Arithmetic aggregation with equal weights across subdomains and domains}

The final step is the arithmetic aggregation of the index, where there are equal weights for subdomains $w_s$ and domains $w_d$, while indicator weights are derived from a factor analysis.  All the weights have been chosen as positive and summing to one, for all the different aggregation levels. 
The Health Index, at the hierarchical levels of indicators, subdomains, domains and overall, is then computed for each year at geographical levels of UTLAs, regions and the nation, where the geographical aggregations at the regional and national levels are population-weighted.

\subsection{The Health Index ranking distribution} 

\begin{figure}[ht]
\centering
\makebox{\includegraphics[width=\textwidth]{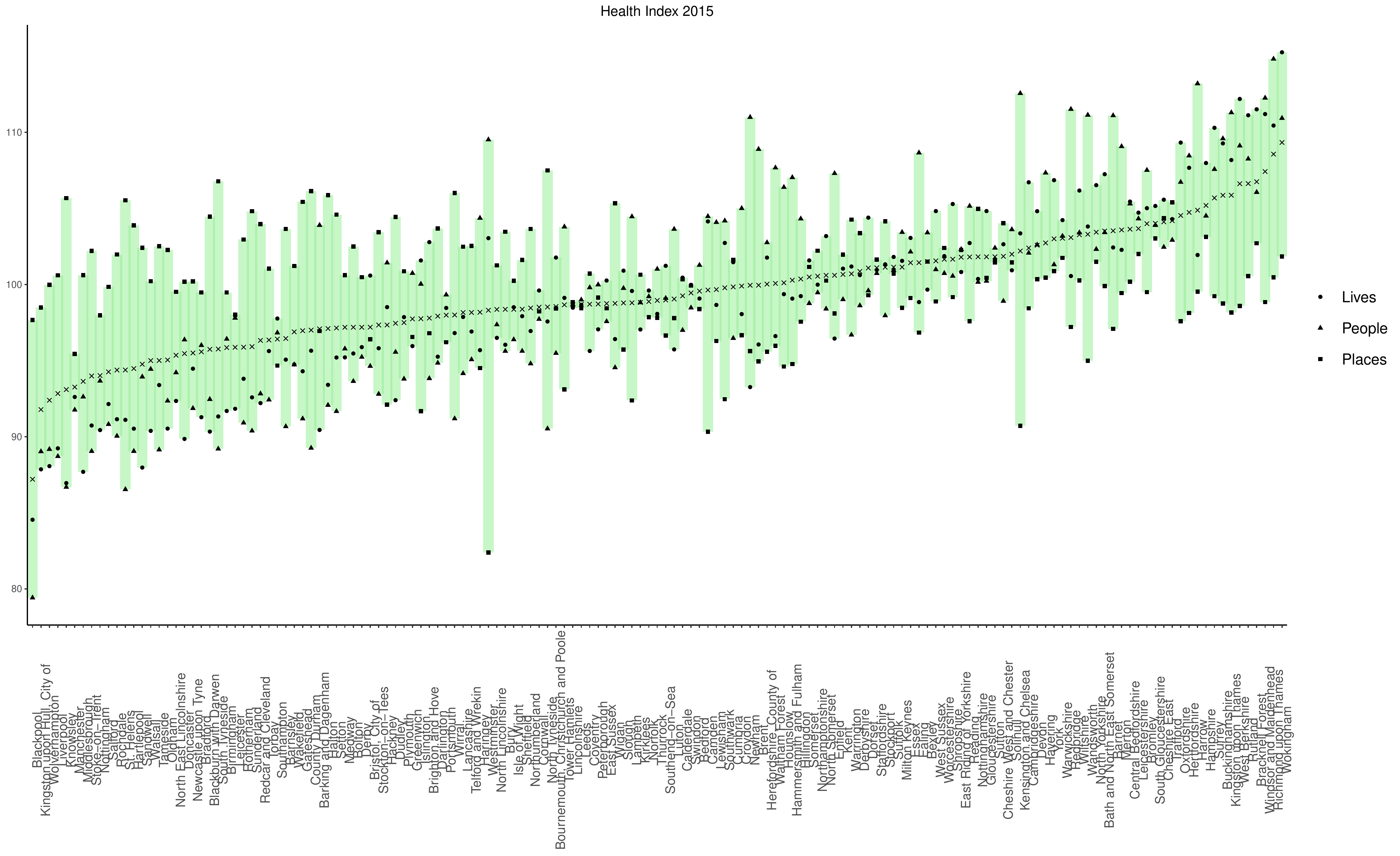}}
\caption{\label{fig:fig1}The 2015 Health Index ordered by UTLA ranking,  jointly with Healthy Lives, Healthy People and Healthy Places indexes, and green bars indicating the minimum and maximum value of the  domains.}
\end{figure}

For the year 2015, for each UTLA, we plot each domain's Health Index values, ordering the UTLAs by the overall Health Index ranking, in Figure~\ref{fig:fig1}. It emerges that Lincolnshire, Leeds and Staffordshire have all three domain index values concentrated at the same values. In contrast, Westminster (the UTLA with the largest difference in domain indexes) presents Healthy People at 109, similar to Kensington and Chelsea, but Healthy Places at 82.  Westminster and Blackpool present similar values for Healthy Places and Healthy People, but their ranking is significantly different.  It is interesting to note that Healthy Lives sits within the range defined by Healthy Places and Healthy People. Similar patterns are observed for the following years, as reported in the SM (see Figures 5-7).

\subsection{A modified ONS Health Index}
\label{sec:modhi}

Before  investigating the HI and carrying out further analysis: correlation and sensitivity/robustness analysis, we implemented a slight change to the original HI as presented above.
As pointed out in~\cite{joint2008handbook}, a certain coherence in the methods needs to be preserved to create a statistically sound index.  
This change was done to avoid statistical misinterpretation, as not all the potential combinations of data transformation and subsequent data operations could be properly interpreted, as carried out in the ONS version. 

Hence, we have computed a modified ONS HI version. We begin from the imputed matrices $\mm X_i$ for each indicator, $i$. Then, instead of directly selecting and applying transformations $f_i$ to ensure normality,  we accounted for kurtosis and skewness using winsorization first and  then by transforming. This approach resulted in only 5-7 variables per year that have been log-transformed (Table  4 in SM). We proceed to standardize using a z-score (following the ONS), and then aggregated with arithmetic mean and equal weights (see Table 5 in SM for comparison ). 

We opted for less strict data transformation, as this  would have not changed the aggregation formula interpretation.
As it stands at the moment, the ONS data transformations included  40 indicators,  with 18 indicators log-transformed. By aggregating all transformed variables using an arithmetic mean, the untransformed variables are effectively aggregated via a mix between a geometric mean (for log-transformed variables) and arithmetic mean (for other variables). 
As succinctly summarised by \cite{nardo2005tools}, ``\emph{when the weighted variables in a linear aggregation are expressed in logarithms, this is equivalent to the geometric aggregation of the variables without logarithms. The ratio between two weights indicates the percentage improvement in one indicator that would compensate for a one percentage point decline in another indicator. This transformation leads to attributing higher weight for a one unit improvement starting from a low level of performance, compared to an identical improvement starting from a high level of performance.}'' 

We used this modified version as the starting point for the rest of this paper. 
The z-scores (see Figure 2 in SM) comparison between this modified version and the original ONS shows that several indicators have more outliers below the 25\textsuperscript{th} percentile, but overall, there are no major discrepancies in values.
Indeed, this modified version generated a different ranking, that affected the UTLAs in the middle, while the top and bottom UTLAs remain unaffected (see Table 4 in SM).
The biggest shift in ranking is observed for Barking and Dagenham (which moved positively 49 positions), whereas Westminster, Herefordshire and  Shropshire all shifted down the rankings by, respectively, 50, 48 and 51 positions. Overall 52\% of the UTLAs shifted in absolute value of within 10 ranking positions, 38\% shifted between 20--30 positions and only 9\% shifted more than 31 positions. Only Blackpool, Kingston upon Hull, City of
Northampton and Hertfordshire kept the same ranking in comparison to the original ONS HI version. 
All the analysis was conducted on R version 4.2\cite{Rcrac} and COINr package \cite{COINr}, by Becker.

\subsection{Proposal}

\begin{itemize}
    \item We suggest to adopt  winsorization, as it takes care of the outliers and provides a robust approach to ensure that kurtosis and skewness are within the acceptable limits, without heavy mathematical data transformations. This also preserves the statistical coherence at aggregation level and interpretation.
\end{itemize}

\section{Correlation analysis}
\label{sec:corr}

The core of every composite index is the indicators, which have to be selected carefully to represent the dimensions of the phenomenon that we are trying to summarize. 
Hence, correlation analysis plays a dual crucial role in the composite indicator construction.  First, statistical analyses anchored on the correlation - such as principal components analysis, factor analysis, Cronbach's alpha - are all suitable to assess that the selected indicators are appropriately representing  the statistical dimensions, i.e. theoretical constructs are supported by the data. Second,  it is useful to identify highly correlated indicators (subdomains and domains), to highlight data redundancy and  potential structure issues.  

Ideally each indicator  (this is  true also for subdomains and domains) should be positively moderately correlated with the others, while high inter-correlations may indicate a multi-collinearity problem and collinear terms should be combined or otherwise eliminated.  
Negative correlations are an undesirable feature in CI, however they may occur at different hierarchical levels of the index. For example, if an indicator is negatively correlated, it can be removed. If domains or subdomains show negative correlation then aggregation by geometric or arithmetic mean should be discarded as it would insert an element of trade-off where units that perform well in one domain have their overall performance  affected by the poor performance on another domain.
To explain how negative correlations affect the composite index, Saisana et al. \cite{saisana2012sustainable} reviewed the  Sustainable Society Index (SSI). The index - similarly composed to the HI -  has  three  main domains: Human, Environmental and Economic wellbeing.  Human and Environmental wellbeing show negative correlation, as in many countries Human and Economic wellbeing  go hand in hand, at  the expenses of the Environment. Their review suggested that  these correlations are a sign of a trade-off, whereby many countries that have poor performance on  Environment levels, have good performance on all other categories and vice versa, therefore each domain should be presented as itself in scoreboard and not aggregated.
This is what happens to Blackpool and Westminster in Figure \ref{fig:fig2}, where Westminster presents the lowest Places indicator and Blackpool  for People, but not for Places. 
We will explore further the trade-off and correlation  and their role in weights definition, but before we provide an extended  HI correlation analyses.

\subsection{Health Index Correlation analysis  } 

\begin{figure}
   \makebox{\includegraphics[width=\textwidth]{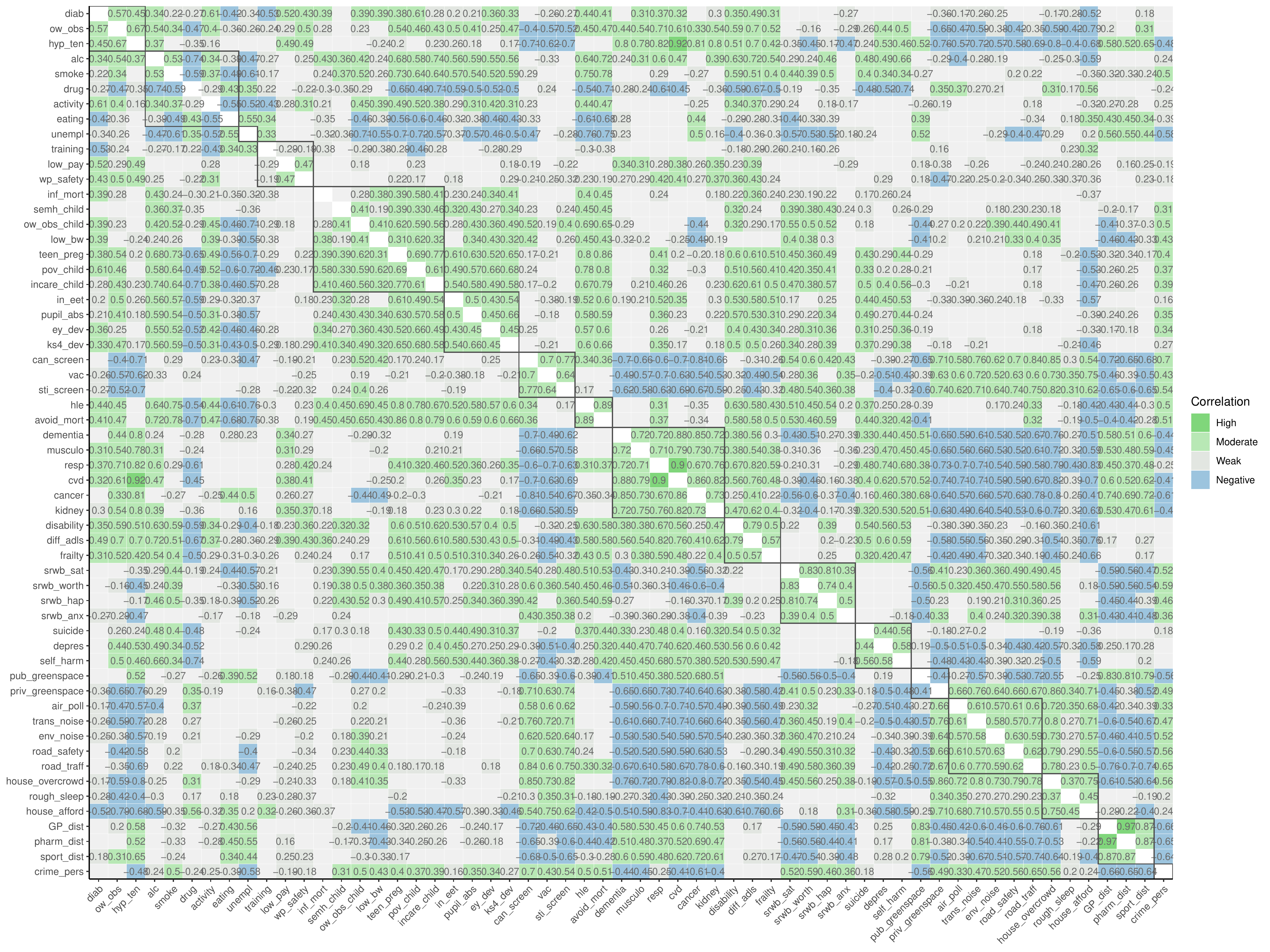}}
    \caption{ \label{fig:Corr1}Correlation heatmap for indicators grouped in subdomains}
   \end{figure}

We used our modified version of the Health Index to carry out a correlation analysis (Pearson) at the different levels of aggregation. The correlation analysis provides insights on the potential redundancy of those indicators with high correlation ($\rho \geq 0.9$); negative correlation ($\rho \leq -0.4$) also indicates some conceptual problems. Acceptable correlation values are for weak ($ 0.3 < \rho \leq 0.4$) and moderate ($ 0.3 \leq \rho < 0.9$).
 The ideal situation would be to have indicators positively correlated among them ( 0.3- 0.9), and not highly correlated with other subdomains as this could impact weights and aggregation. 
 In Figure \ref{fig:Corr1}, indicators grouped in subdomains are showing  overall positive correlations. However, there are some correlations of concern. For example, public and private green space that define the subdomain 'Access to green space (Pl.1)' show negative correlations, and in subdomain  'Access to services (Pl.4)'  distance to the nearest pharmacy and general practitioner (GP)   are also highly correlated. We suspected that this could be somehow related to the urban/rural UTLA definition. 
Cardiovascular and respiratory  prevalence are highly correlated in subdomain 'Physical health conditions (Pe.2)'. 
The indicators in  'Behavioural risk factors (Li.2)' and 'Working conditions (Li.4)'   present negative  and weak correlations.
In this heatmap, we see also correlation among the subdomains like blocks. For example 'Risk factors for children(Li.5)' and  'Children and young's people education (Li.6)'  are also correlated, likewise 'Physical health conditions (Pe.2)' and 'Difficulties in daily life (Pe.3)'.

\begin{figure}
     \makebox{\includegraphics[width=\textwidth]{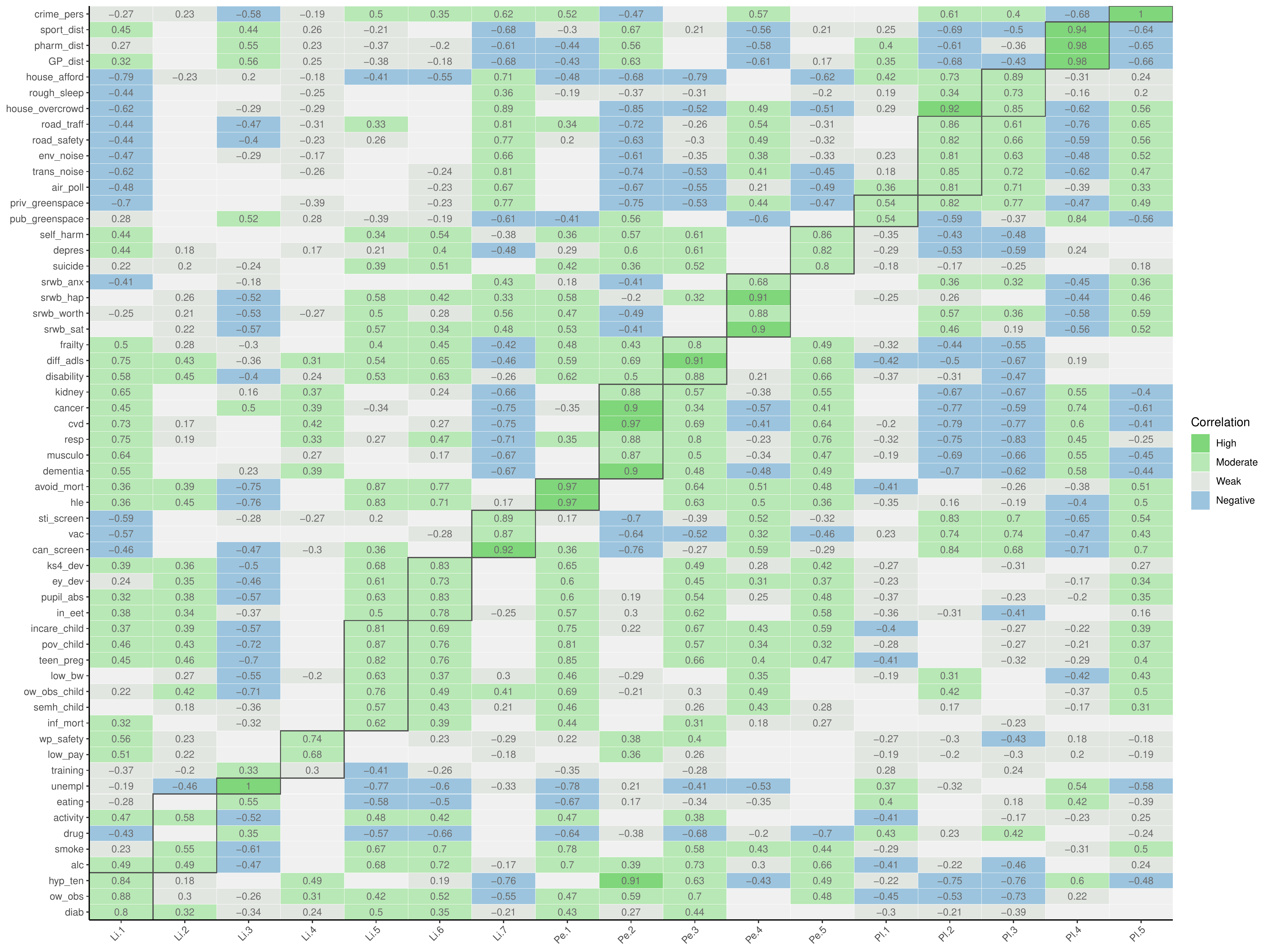}}
    \caption{ \label{fig:SubCorr1}Correlation heatmap for indicators  and subdomains, grouped by subdomains  mean.}
   \end{figure}

From the subdomain correlation map (see Figure \ref{fig:SubCorr1}), we immediately see  that the indicator 'Household overcrowding' is highly correlated with the subdomains on 'Local environment (Pl.2)'.
Finally, we correlated ( see Figure  3 in SM) subdomains versus domains, we found that  People subdomains are overall well correlated with the other subdomains within their domain. Lives and Places are similar but present some weak correlations: 'Access to services (Pl.4)', and  'Unemployment (L1.3)' and 'Difficulties in daily life (Pe.3)'. This confirms what we have observed in the indicators heatmap. 

 \begin{figure}[ht]
 \makebox{\includegraphics[width=\textwidth]{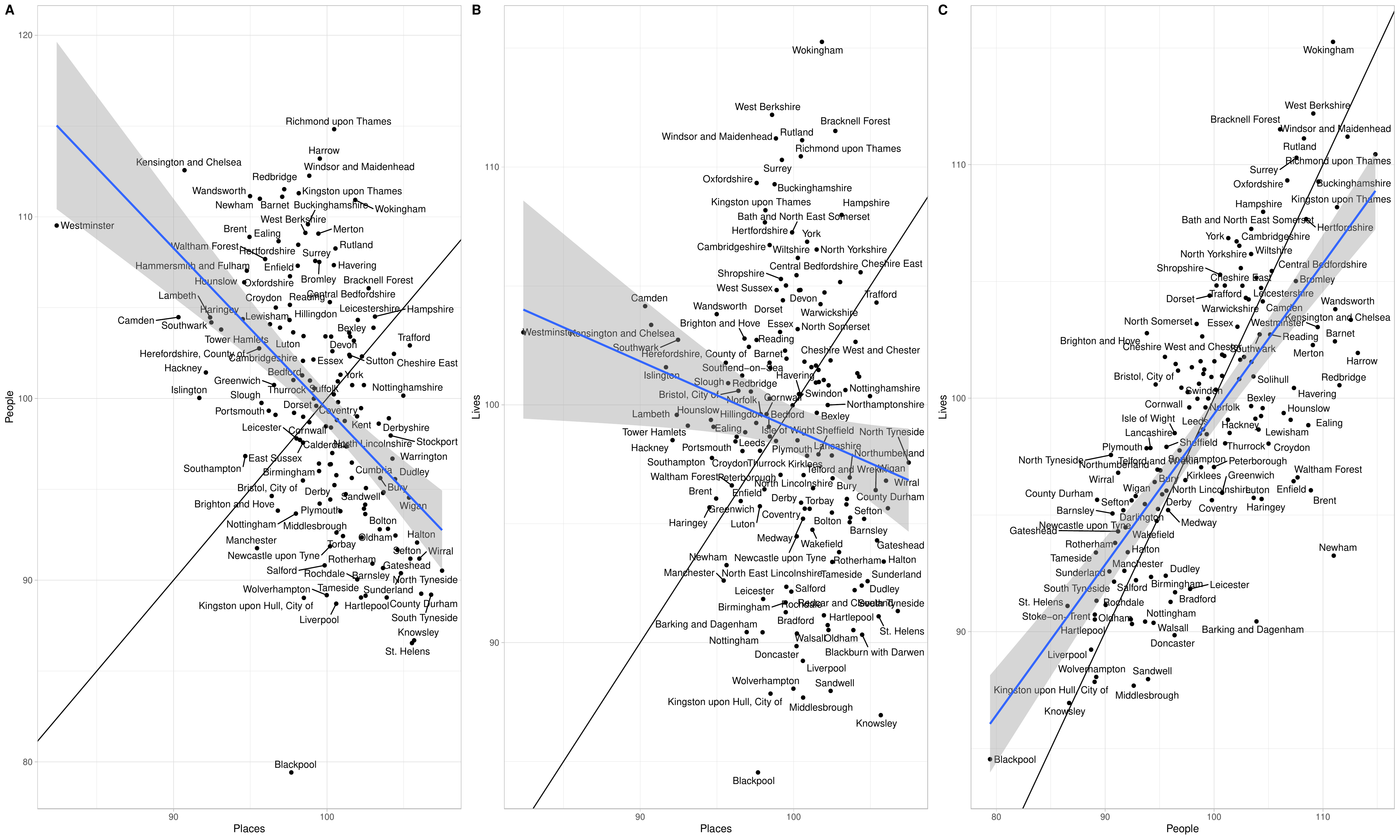}}
 \caption{ \label{fig:fig2} UTLA Domains index scatter plots with fitted linear regression (2015): (A) People vs Places, (B) Lives vs Places, (C) Lives vs People.}
\end{figure}

The panels in Figure~\ref{fig:fig2} show the scatter-plots for the three domains. It can be observed that 
Healthy Lives and Healthy People have a high Pearson correlation ($\rho$= 0.65), while for Healthy Lives and Healthy Places ($\rho$= -0.12) and Healthy People and Healthy Places ($\rho$= -0.39) the correlations are negative, a similar situation as described for the SSI by \cite{saisana2012sustainable}. Once we removed London's UTLAs, characterized by high values of People and Lives and low on Places, the correlation for  Lives and  People increases ($\rho$= 0.72),  null for  Lives and Places ($\rho$= -0.06) and diminishes in People and Places  ($\rho$= -0.25).

\subsection{Proposal} 

\begin{itemize}
    \item We suggest to  remove the public and private green space indicators due to the high negative correlation.
    \item A revision of the indicators defining 'Behavioural risk factors'. Physical activity is correlated with alcohol misuse and  smoking and not with healthy eating, which is correlated with drug use. The subdomain should be  split and re-organized. Drug misuse and healthy eating  are pointing  toward unemployment and in general to a  some measure on society inequality. 
    \item Subdomains 'Risk factors for children' and 'Children and young people’s education' could be merged in a unique block, as the indicators are highly correlated. 
    \item Cardiovascular, respiratory and  hypertension could be combined as they are highly correlated, therefore bringing redundancy. 
\end{itemize}

\section{The choice of a weight system}
\label{sec:weights}

In this section,  we introduce the choice of a weights system that could be  employed in the linear aggregation formula that generate the composite index.  We review the definitions and how the weights can be interpreted. We then proceed on evaluating what role this plays for the correlation at  different levels (indicators/subdomains/domains) and we describe the optimized method~\cite{becker2017weights} that generates weights that account for correlations. 

We also compared the time-series factor analysis derived weights, currently in use in the ONS HI  with that for the ONS HI with weights generated by principal component analysis (PCA). 
We introduce them here, because we are going to use the PCA weights and the optimized weights as options in our sensitivity and uncertainty analysis. 

\subsection{Weights definitions: compensatory versus non-compensatory}

In standard practice~\cite{COIN,munda2005constructing}, the composite indicator for time $t$ is defined as:
\[
z_{t} = \sum_{c \in C} w_{ct} z_{ct},
\]
where $c$ indexes the indicators and $C$ is a set of indicators being composed (which, in the context of the ONS HI, may correspond to subdomains, domains, or the overall index).
Thus the composite indicator is a weighted linear aggregation, where weights are (typically) constrained to sum to 1. 

In the composite index literature \cite{greco2019methodological}, weights methods are often found to be linear, geometric or multi criteria, or classified into compensatory and non-compensatory approaches.
However, the major difference in weight systems boils  down to  defining weights either as coefficients that address the importance of a variable (indicator/subdomains/domains) or as a trade-off coefficient.  

Weights that convey `importance' should be used in aggregation formulae that do not allow for compensability; that is, where poor performance in some indicators can be compensated by sufficiently high values of other indicators.  These  definitions are also known as compensatory, because the  `compensation' refers to a willingness to allow high performance
on one  variable (subdomain/domain) to compensate for low performance on another. The weighted mean (arithmetic/geometric)  is a classic example of compensatory approach, where the weight is a de facto  trade-off coefficient. 

Non-compensatory methods  allow the  weights to express `importance', where  the greatest weight is placed on the most
important `dimension' \cite{vincke1992multicriteria, vansnick1990measurement}. These approaches have their roots in social choice theory (also known as multicriteria) and more details can be found in~\cite{munda2008social}. Briefly, in this framework,
indicator (subdomain/domain) values rank  the countries in  different ways and contributed to  define the relative performance of each country/option with respect to each of the other countries/option. This indicators-unit ranking, generates an impact matrix and a voting system must be put in place to define the overall ranking. For example  the `plurality vote' will rank as first the unit (UTLA) that has ranked at first place on the majority of the indicators.  However, this approach comes with the price of dealing with preferences and choices  on how to select the final ranking given the indicators-ranking \cite{munda2008social}. Two popular  approaches, that take the name after their authors,   suggest that a Condorcet approach is necessary when weights are to be understood as importance coefficients, while a Borda approach is desirable when weights are meaningful in the form of trade-offs. 

These methods, while valuable, are rather harder to implement as they require an expert panel to grade the indicators in first place, but also lack the immediate facility to explain when compared with a weighted mean. The dual notions of weighting as importance versus weighting as trade-off and their interpretation requires  more consideration, to assure that selected weights are in line with the practitioner preferences.  In their article, Munda and Nardo~\cite{munda2009noncompensatory, munda2005constructing}  provide extensive commentary on  an interesting mis-interpretation around the weights/aggregation combination that gets buried in the CI construction, but is useful to address here.

 \subsection{Weights as 'importance' coefficients, linear aggregation and correlation} 
 
According to the  OECD  guidelines~\cite{freudenberg2003composite}: ``Greater weight should be given to components which are considered to be more significant in the context of the particular composite''.
As pointed out by Greco et al.~\cite{greco2019methodological}, the popular linear aggregation weights are used as if they were
importance coefficients, while they are in fact trade-off coefficients. 

Briefly, the authors \cite{munda2005constructing, munda2009noncompensatory} state that in linear and geometric aggregation the weights play the role of a trade-off ratio that depends on the scale of measurement.
If the weight has to be interpreted as a measure of importance, then the weights should be connected with the indicators themselves and not with their quantification; they should be invariant to the units of the indicator. 
This distinction between weights as trade-off ratio versus  importance does not disappear even when all indicators are on the same scale. For a weight to express `importance', then non-compensablity should be enforced. This issue  becomes relevant when CI are composed of different data for multicriteria optimization   where improvement in one domain cannot compensate for degradation in another. One way to disentangle this paradox of trade-off weights interpreted as importance weights is proposed by Becker \cite{becker2017weights}. In order to derive  weights as `explicit importance', we need to evaluate the correlation structure  and use it to understand the `importance' role of the domains/ subdomains in the composite indicators, and what the influence of each indicator is on the index, generating optimized weights.

\subsection{Optimized compensatory weights}

For a weighting system where  weights are representing `explicit importance', then different variances and correlations among indicators (subdomains/domains) mask the weights to represent importance, as shown above in the correlation analysis.  

To find weights that reflect importance and not trade-off ratios, conditioned on the correlations, we follow the methods introduced by Becker \cite{becker2017weights}.
We recall that  for the Health Index, domains and subdomains have equal weights, while indicators have data-driven  weights derived by FA. If we take the equal weights choice as a way to express equal importance of the three domains,  we need to account for each variable's influence on the output,  and how weights can be assigned to reflect the desired importance, `conditioned' on the existing shared information among the domains. Knowing the correlation among domains can help to reduce uncertainty, as strong correlations suggest that the domains should be treated jointly, rather than individually.
This can help in reassessing the weights.

A measure of importance, capturing the dependence between the CI and the effect of  domains, starts from analysing the correlations ratio $S_d$, also known as the first order sensitivity index or main effect. We split the correlation ratio in two parts: a correlated part, $S_d^{c}$, and uncorrelated part, $S_d^{u}$, such that
\[
S_d=S_d^{c} + S_d^{u}
\]
where $d =1,2,3$ indicates the level of aggregation. 

A large value for $S_d$, with a relatively low uncorrelated part $S_d^{u}$ such that $S_d \approx S_d^{c}$, indicates that the domain contribution to the index variance is only due to the correlation with the other domains~\cite{mara2012variance}. However, if $S_d^{c}$ is negative, this implies conceptual problems with one of the domains, and is not a desirable feature in composite indicators.

The optimized weights have been presented in Becker et al.~\cite{becker2017weights}.
It is important to note that, while we have applied this approach at the domain level,  the same methodology can be applied to other levels of the hierarchical structure of the Health Index, i.e. for aggregating indicators into subdomains and subdomains into domains.
Briefly, first we estimate $S_d$ and $S_d^{u}$ by implementing a series of linear and non-linear regressions (using splines~\cite{wood2001mgcv}). The steps to compute the two summands of the correlation ratio, are the following:

\begin{enumerate}
\item Estimate $S_i$ using a nonlinear regression approach 
\item Perform a regression of $x_d$ on $x_{\sim d}$. This can be either linear (using multivariate linear regression), or nonlinear (using a multivariate
Gaussian process). Denote this fitted regression as $\hat{x}_d$.
\item Get the residuals of this regression, $\hat{z}_d=x_d- \hat{x}_d$.
\item Estimate $S_d^u$ by a nonlinear regression of $y$ on $\hat{z}_d$, using the same approach as in step (a).
\item The correlated part then is  the simple expression $S_d^c=S_d-S_d^u$
\end{enumerate}

Using a simple numerical approach, the weights are estimated that result in the desired importance, using an optimisation algorithm.
If $\tilde{S}_d=\frac{S_d}{\sum_{d=1}^D S_d}$  is the normalised correlation of $x_d$, then the targeted normalised correlation ratio is $\tilde{S_d^\star}$, where it is assumed that is 
$\tilde{S_d^\star}=w_d$ is the weight assumed (in our case equal weights)
to reflect the importance. 
Once these quantities have been computed, and provided the equal weights ( or any other weight system user-provided), the optimized weights are the results of the minimizing the objective function
\[
w_{opt}=\sum_{d=1}^D ( \tilde{S}_d^\star-\tilde{S}_d(w) )^2.
\]

\begin{figure}
\centering
\makebox{\includegraphics[scale=0.5]{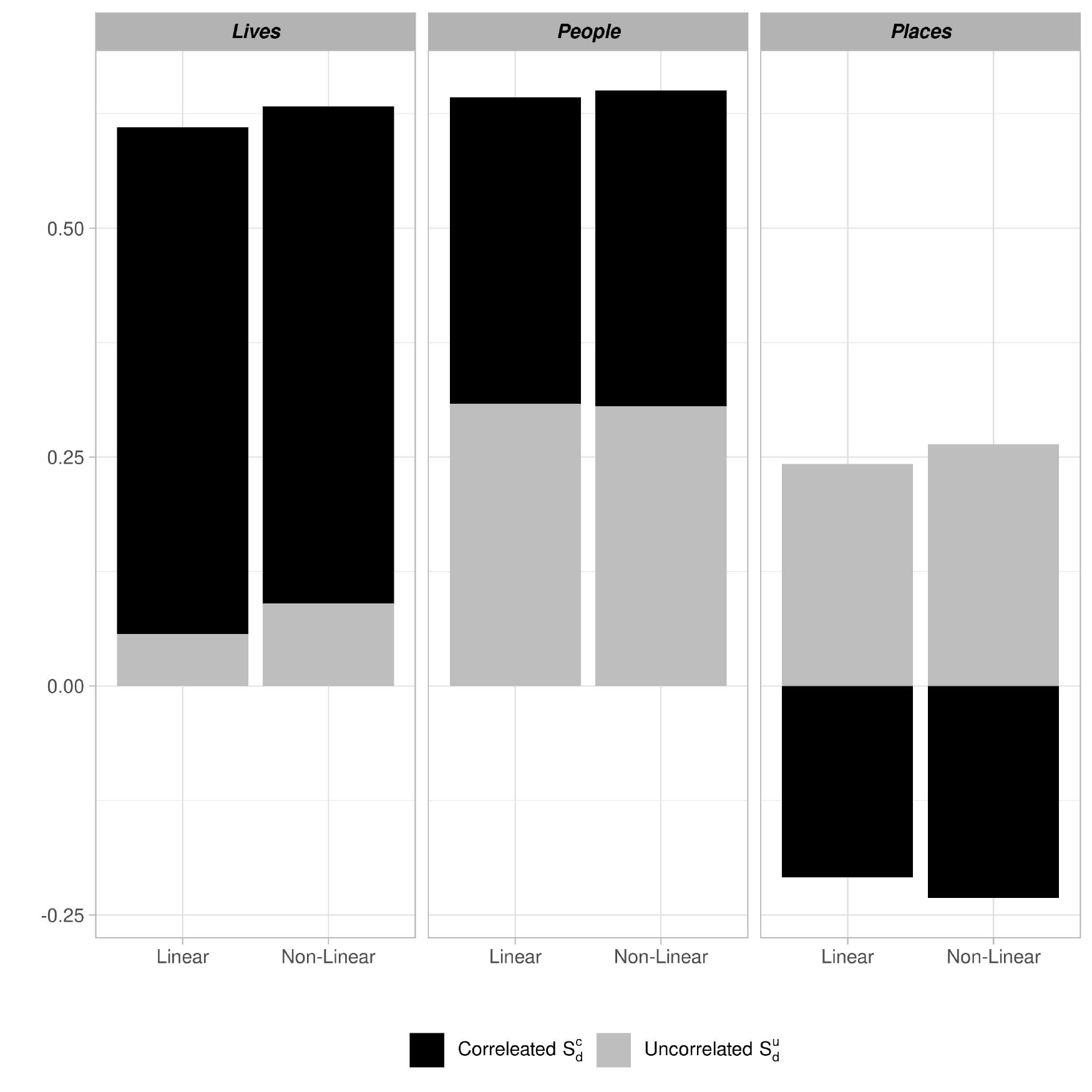}}
\caption{\label{fig:fig4}Estimates of $S_d$ (full bars), broken down into correlated $S_d^c$ and uncorrelated $S_d^u$, using linear and non-linear dependence modelling.}
\end{figure}

The results of this approach are reported in Figure~\ref{fig:fig4}. 
Our first observation is that the domains have fairly similar linear and non-linear  correlation ratio $S_d$ estimates, indicating that linear estimates would have been sufficient to address the linear correlation among the three domains.  
Recall that we would like low $S_d^c$ and high $S_d^u$, both positive. 
What we have obtained is that the correlated part dominates in Healthy Lives, indicating that Healthy Lives has a small impact on the composite index as it is mostly imputable to the correlation with the other variable. 
For Healthy People both components contribute equally and both positively. 
Healthy Places has a negative correlated effect, which  is similar to the uncorrelated part, but the negative $S_d^c$ values implies potential problems in the composite index. Somehow, we could have expected that Healthy  Places could  have some problematic behaviour, as we have observed in the correlation analysis. 

Having unpacked the correlation among the domains, we can use this information to find a new set of weights that truly reflects the importance of each variable in the CI, but that are close to the importance distribution we have specified - in our case equal importance (each domain 0.33 weight). The optimization algorithm finds optimal weights of 0.45  for Healthy People, 0.16 for Healthy Lives, and 0.73 for Healthy Places.
These weights will be used subsequently for a sensitivity and uncertainty analysis.

\subsection{Principal Component Analysis derived weights}

While there is no objective choice in selecting the weights, we concentrate on a so-called data-driven weighting system, derived from Principal Component Analysis (PCA) or Factor Analysis (FA).
Now, in the context of composite indicator construction, these two methods can be applied at different steps due to their versatile interpretation: to identify dimensions, to cluster indicators and to define weights. While PCA and FA share several methodological aspects, there is a key difference between the two analyses. PCA is a data reduction method based on the correlation matrix, which re-defines a new set of uncorrelated variables as linear combinations of the original variables. In contrast, FA is a measurement model of a latent variable, where the latent factor 'causes' the observed variables. 
There is a \emph{recommendation} in the CI community~\cite{saisana2002state} to use the PCA loadings as weights only if the first component accounts at least for the $70\%$ of the total variability.
We applied this procedure to derive the weighting systems for subdomains. 
The 58 indicators are split in 17 subdomains (see Table~1), and for each of these subdomains we carried out a PCA analysis, for each year.

For most subdomains, over all four years, the first PCA component accounted for a range between 51\% to 94\% of the total variability. Exceptions were observed (see Table  6 in SM) for 'Mental health (Pe.4)' with variance explained 66-69\%, 'Behavioural risk factors (Li.1)'  53-55\%, 'Working conditions (Li.4)' 50-55\%, 'Risk factors for children (Li.5)' 55\%, 'Children and young people’s education (Li.6)' 63-69\% , 'Access to housing (Pl.3)' 65-69\%.
We then normalized the loading coefficient and compared them over time, jointly with the weights originally derived from FA for all the years collapsed. 

\begin{figure}[ht]
\centering
\makebox{\includegraphics[width=\textwidth]{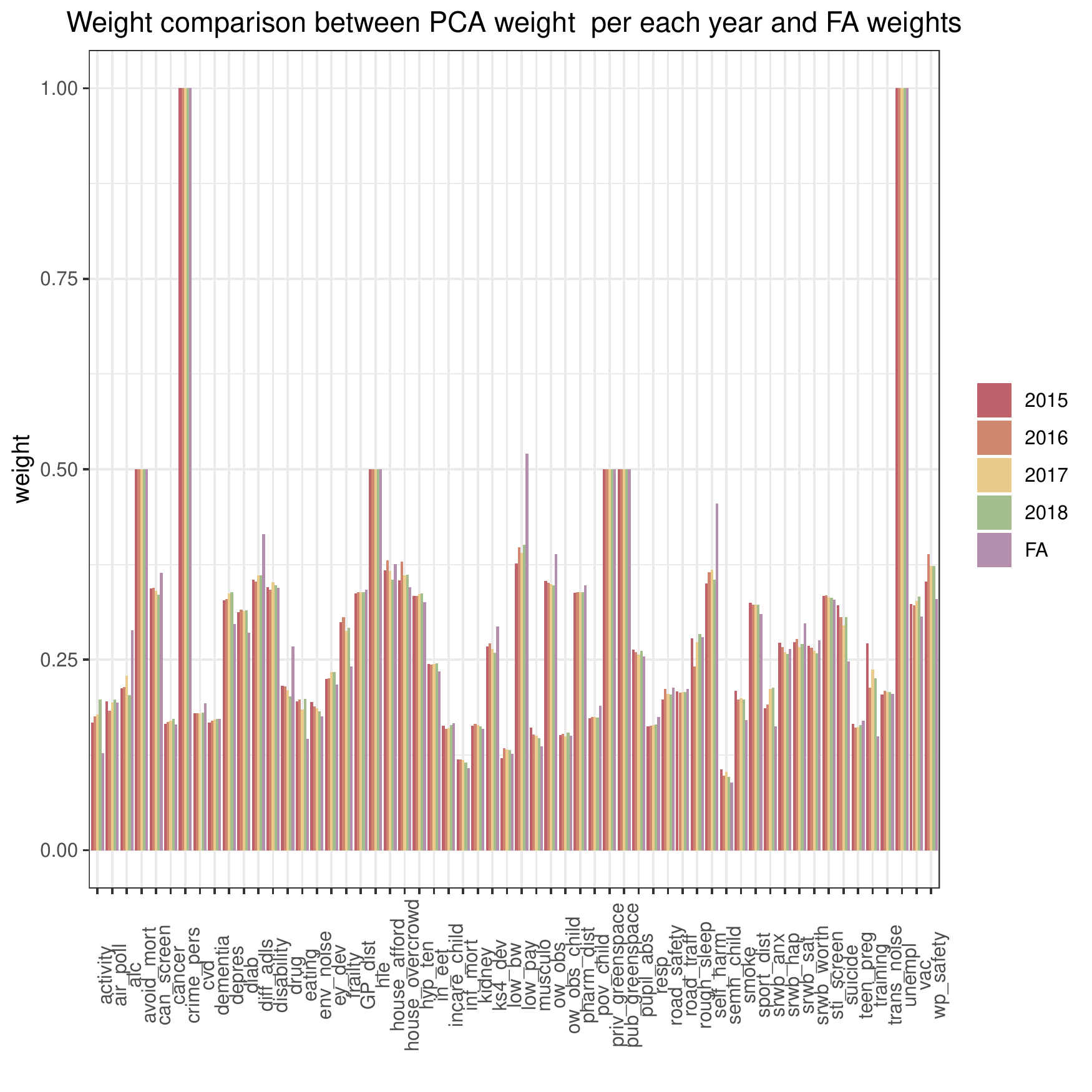}}
\caption{\label{fig:fig3} Weight comparison between PCA weight  per each year and time-series FA weights.  [labels need to be changed to more readable]}
\end{figure}

We have investigated the PCA weights values over time and compared them with the time-series  FA analysis computed for the ONS HI.
We have found that these are very similar over time, which is reassuring in terms of stability of the index weights (see Figure \ref{fig:fig3}). However, when we compared PCA and FA weights, we have  found that FA gave higher weights to the following indicators (difference percentage among weights): low pay (12\%), self-harm (10\%), difficulty completing
activities of daily living (5.4\%) and drug misuse (6.6\%).
On the contrary, PCA imposed higher weights to job-related training (7.6\%), physical activity (7\%), suicides (5.9\%) and workplace safety (4.4\%).

\subsection{Proposal}

\begin{itemize}
    \item Given the negative correlation between the Healthy Places domain and the two other domains, we discourage arithmetic mean aggregation.  
    \item To interpret the weights as `importance', optimized weights should be adopted. 
    \item Between FA and PCA derived weights, we recommend using PCA as the subdomain is not the `cause' of the indicators, but a combination. 
    \item The overall index could be obtained by adopting optimized weights and a geometric mean aggregation formula. Optimized weight  allow for partial substitution as correlation is accounted for; geometric mean rewards balance by penalizing
    uneven performance in the underlying domains. 
\end{itemize}

\section{Sensitivity and Uncertainty analysis}
\label{sec:sens}

Following the approach introduced by Saisana et al.~\cite{saisana2005uncertainty,sobol1993sensitivity,saltelli2010variance},  we carried out an analysis of the sensitivity and uncertainty of the Health Index.
This analysis is based on a variance-based approach that constructs Monte Carlo estimates of the variability observed due to each step, and due to the interactions between the different steps.
For each of the construction steps $\mathbf q_{i}$  we select  a potential  alternative  methods. Therefore, indicating the model with $m$, we can compute the global variance as
\[
  V(m)=\sum_i V_i+\sum_i\sum_{j>i}V_{i,j}+...+V_{1,2,...,k},
\]
where

 \begin{align*}
 V_i =& V_{q_i}[E_{\mathbf q_-i}(m|\mathbf q_i)],\\
 {V_i}_j =& {V_q}_{ij}[E_{\mathbf q_-ij}(m|\mathbf q_i,\mathbf q_j)] - 
V_{q_i}[E_{\mathbf q_-i}(m|\mathbf q_i)]-
V_{q_j}[E_{\mathbf q_-j}(m|\mathbf q_j)]
 \end{align*}

The quantity $V_{q_i}[E_{\mathbf q_-i}(m|\mathbf q_i)]$  and the  expectation $E_{\mathbf q_-i}$ require the computation of an integral
 over all factors except $q_i$, including the marginal distributions for these factors. The variance $V_{q_i}$ would imply a further integral over $q_i$ and its marginal distribution.

The sensitivity indices are then  $S_i=V_i/V(m)$.
These terms measure the contribution of the input $\mathbf q_i$ to the total variance, and can be interpreted as a fraction of uncertainty. 
  
The first order sensitivity index, which is the fraction of the output variance caused by each uncertain input assumption alone,  is: 

\[
S_i = \frac{V[E(m|\mathbf q_i)]}{V(m)},
\]
this is averaged over variations in other input assumptions, and the
total order sensitivity index, (or  interaction),
\[
S_{Ti} = 1 - \frac {V[E\left(m \mid \mathbf q_{-i} \right)]}{V(m)} = \frac {E[V\left(m \mid
    \textbf{q}_{-i} \right)]}{V(m)}
\]
where $\mathbf q_{-i}$  is the set of all uncertain inputs except the $i$\textsuperscript{th} quantity, and the quantity $S_{Ti}$ measures the fraction of the output variance caused by  
$\mathbf{q}_{i}$ and any interactions with other assumptions.
In carrying out  the  sensitivity analysis, we have selected potential steps $\mathbf{q}_{i}$  that are coherent with a final linear aggregation.

\begin{table}
\caption{\label{tab:tabsa0} Steps and methods used in the sensitivity analysis}
\fbox{%
\begin{tabular}{*{2}{c}}
\centering
Steps & Alternatives \\
\hline
Data treatment &  winsorization \\
            & (2$^{nd}$,5$^{th}$,10$^{th}$ points)\\
Normalization & z-score, min-max\\
Weights Indicators & equal weights, \\
           & principal components weights \\ 
Weights Domains & optimized weight\\
\end{tabular}} 
\end{table}

The steps and the methods to be tested are listed in Table~\ref{tab:tabsa0}.
In our analysis we evaluated (for 2015) the following main outcomes:
UTLA ranking by overall Index value and 
UTLA rankings by each domain's index value.

We opted for winsorization to control data kurtosis and skewness, by winsorising at the second, fifth and tenth values. We allowed for two normalization  types:  z-score  centered at 100 and standard deviation at 10; and min-max bounded 1-100. For the weights we allowed equal weights, PCA derived and optimized weights for domains only, as previously introduced. We ran the computations for 10,000 iterations. 

We studied also the absolute mean ranking shift of removing indicators and subdomains, to evaluate the roles played by the hierarchical elements.

\subsection{Results for Sensitivity and Uncertainty analysis}
\label{sec:stat}

\begin{figure}
\centering
\makebox{\includegraphics[width=\textwidth]{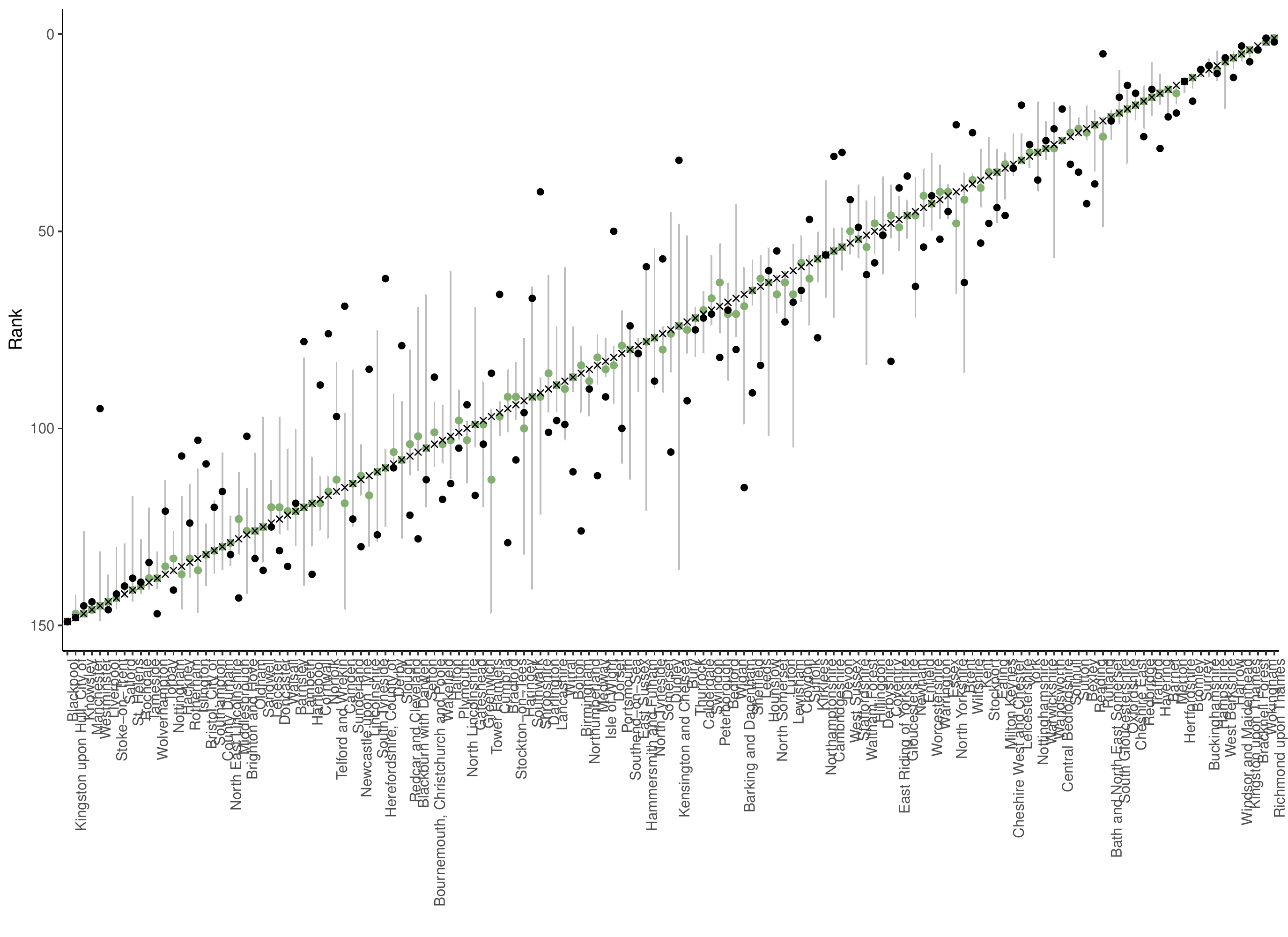}}
\caption{Results of UA showing the overall Index for each UTLA, ordered by the  modified ranking for 2015 (crosses).  With  the corresponding 5\textsuperscript{th} and 95\textsuperscript{th} percentiles (bounds) and the median ranking  (green dots). For comparison,  the original ONS UTLA ranking (black dots).}
\label{fig:figSA1}
\end{figure}

 We carried out the sensitivity analysis on the modified ONS HI and in Figure \ref{fig:figSA1},  we notice that for the overall index tail rankings are stable, while the middle UTLAs are the ones showing the highest variability with median rankings (green dots) above or below the provided ranking.  
 
We  then  repeated the analysis for the three domains separately (see Figure  4 in SM). The estimates are more precise, as the bounds between the 5\textsuperscript{th} and 95\textsuperscript{th} centile are narrower compared to the overall index. We observed that People  rankings are  quite precise and concentrated and  it is possible to see that they are following the Health Index. Lives and  Places are  displaying higher variability,  with  Places acting as the `wild card'.  

\begin{figure}
\centering
\makebox{\includegraphics[width=\textwidth]{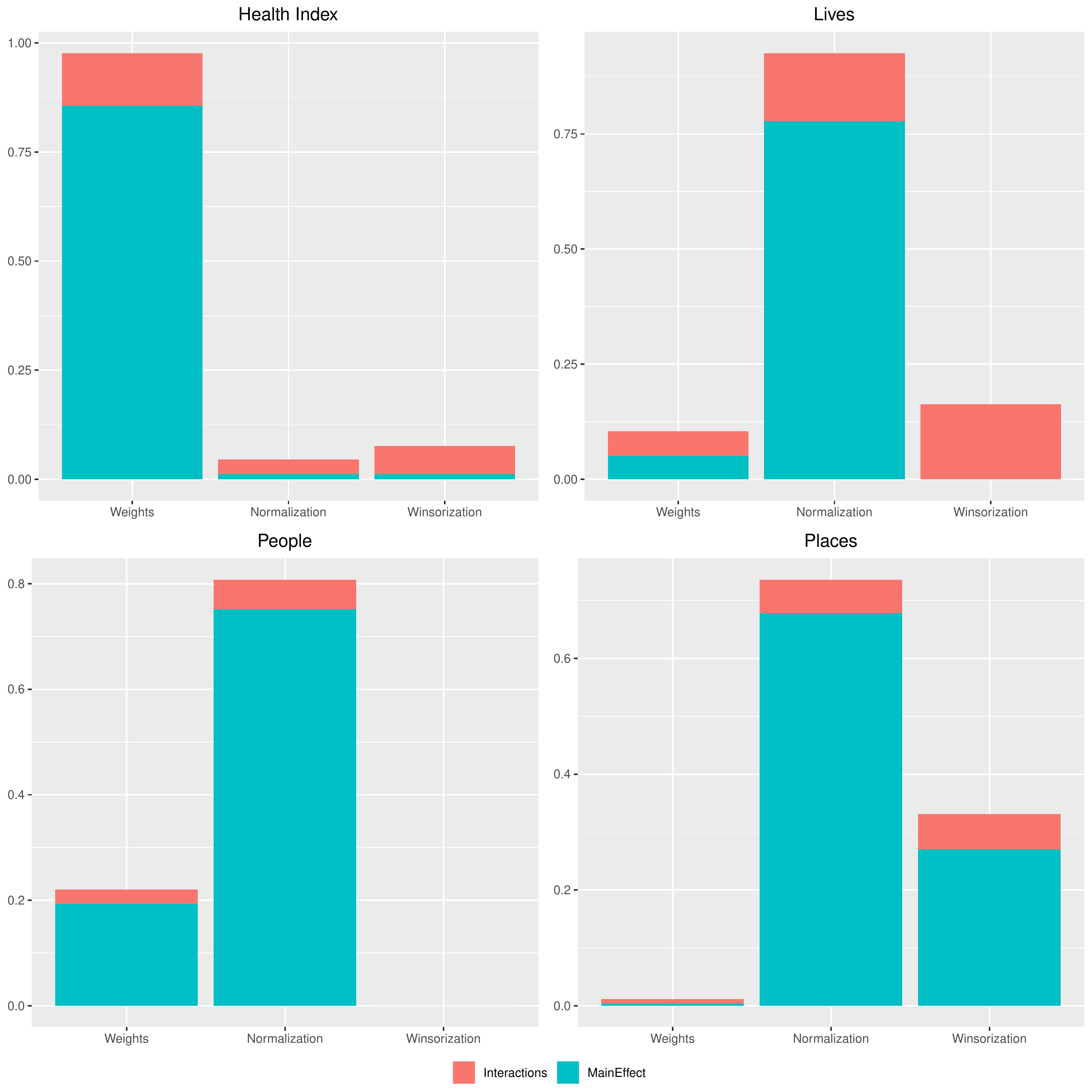}}
\caption{Results of the sensitivity order: Main Effect $S_i$ and the interaction $S_{T_i}$}
\label{fig:figSA2}
\end{figure}

The first order sensitivity and the total order sensitivity have been computed for the overall index and the three domains and we reported them in Table 7 in SM. We then plotted  the  main effect $S_i$ and the interactions $S_{T_i}$, see Figure \ref{fig:figSA2}. These  values  can be interpreted as the uncertainty caused by the effect of the $i$\textsuperscript{th} uncertain parameter/assumption on its own. The total order sensitivity index is the uncertainty caused by the effect of the  $i$\textsuperscript{th} uncertain parameter/assumption, including its interactions with other inputs. This disentanglement shows that at domain level normalization plays a major role for all of them, with winsorization additionally being quite relevant for Places and weights being relevant for People. For the overall index, weights are the  main cause of the variability with normalization and winsorization playing a minor role at interaction levels.

\subsection{Ranking shifts by removing indicators and subdomains}

\begin{figure}
\centering
\makebox{\includegraphics[scale=0.3]{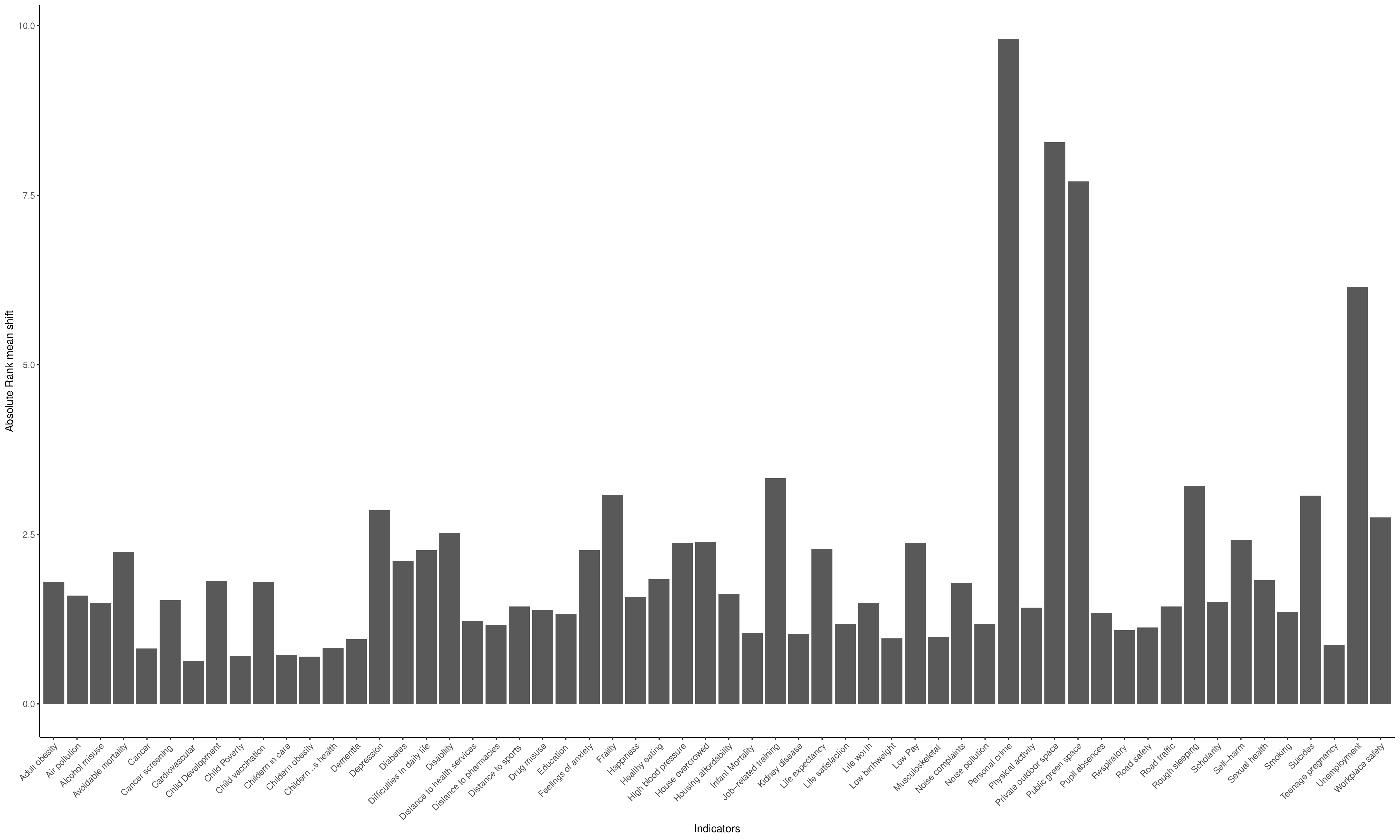}}
\caption{\label{fig:figPillarsInd}The absolute mean rank shift for the overall index by removing one indicator at a time.}
\end{figure}

We assessed the absolute mean differences on the overall rank shift, by removing indicators and subdomains.  At indicator level (see Figure \ref{fig:figPillarsInd}), we observed the highest shifts are due to unemployment, access to private and public  green space and personal crime.  Moderate absolute shifts are observed for job-related training,
workplace safety, disability, frailty, suicides,  depression  and rough sleeping.

\begin{figure}
\centering
\makebox{\includegraphics[scale=0.7]{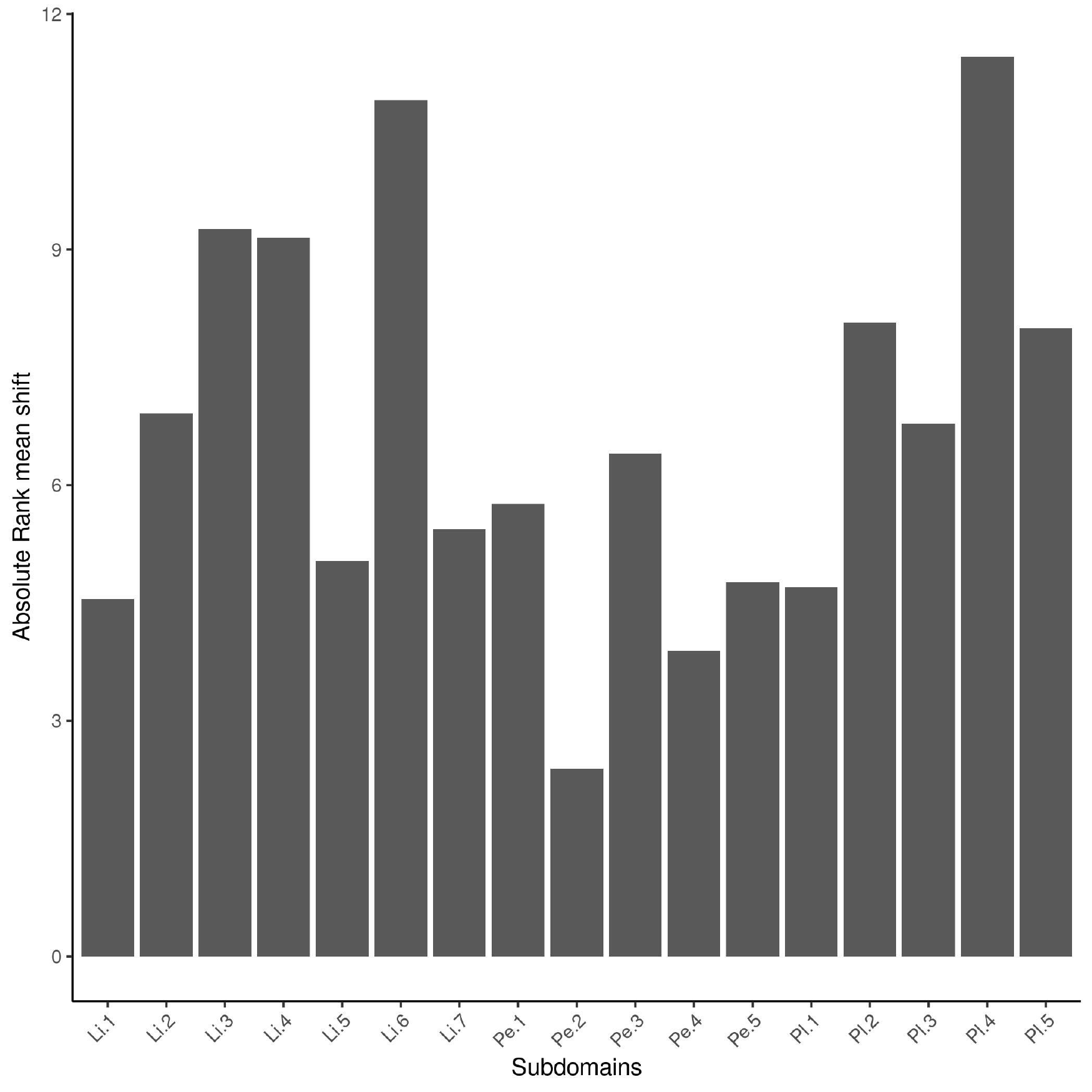}}
\caption{\label{fig:figPillars}The absolute mean rank shift for the overall index by removing one subdomain at a time.}
\end{figure}

At subdomain levels (see Figure \ref{fig:figPillars}), the highest impact is for `Access to services (Pl.4)' and `Children and young people's education (Li.6)', followed by `Unemployment (Li.3)' and `Working conditions (Li.4)'.
The observation that Healthy Lives shows the most influence on the overall index values confirms what has already been observed in previous sections, where we note the high correlation between Healthy Lives values and the overall index values.

\section{Discussion}
\label{sec:conc}

We have scrutinized the choices made when constructing the ONS Health Index for England, and have evaluated the issues that emerge while assessing each construction step. The resulting Health Index is easy to explain to wider audiences, and the data collection and  the index structure are harmonized to be comparable across time and different geographies.  The indicator selection covers the main areas of Health, in line with the WHO definition, and provides access to policy makers to different combination of indicators and comparisons. 

Our analysis  has shown that the weights and normalization steps play a major role in the exhibited variability in the Health Index, in particular for middle-ranking UTLAs. 
The steps that generate the most cumbersome decisions to be taken are the choice of the  weighting system, and the choice of aggregation formula~\cite{greco2019methodological}. However, choices made for both steps need to be taken in the context with the preceding steps. 
Driven by the desideratum to have an index that is easy to explain, we decided to explore in the sensitivity and uncertainty analysis only those methods that were compatible with the approach taken by the ONS. In our case, we considered the use of different weighting systems and data treatment, while staying consistent with a final linear aggregation formula (i.e. an arithmetic mean). This coherence was also the reason why we recommended to intervene  minimally on the data treatment, opting for winsorization and then if still needed we followed with a transformation to normalize the indicator. 
The negative correlation exhibited by Healthy Places, the effect on the rank shifting for Healthy Places indicators, and the low ranking correlation with the overall index, could potentially help us to reflect on the choices of the indicators and potentially revise the indicators selected.
However, it is accurate to claim that areas with worse Healthy Places indicators, such as London boroughs (comprising 20\% of all UTLAs), score higher values on the other two domains.
The reverse is also true, where more rural UTLAs have, for example, lower pollution and good access to private and public green space, but are lower on other indicators. 

By exploring the data derived weights using PCA and comparing them with the initial choices made in the ONS version, we saw some differences, but no major discrepancies. This approach also yielded similar results across time. The fact that PCA and FA return similar results, which are then reflected in weights, could be explained by the fact that, overall, the subdomains are composed of a very limited number of indicators.
Indeed, the highest number of indicators in a given subdomain is for Li.5, with 7 indicators.
Therefore the PCA correlation matrix closely resembles the off-diagonal FA correlation matrix.  

The optimized set of weights allowed us to uncover the relationships among the domains. We could also extend this approach to subdomains.
We have found that the correlation among the domains could be explored by decomposing the correlation ratio in two parts, and that these estimates can be further used to reflect weights as importance and not as trade-off ratios.

The weights play a major role in the sensitivity and uncertainty analysis, while the ranking uncertainty is smaller at people level only.
Once we evaluated the overall index, we observed higher variability for the middle UTLAs. The UTLAs at top and bottom tend to remain stable.
When we compared  the difference between the original ONS  ranking and the rankings range based on the modified index, we found these middle UTLAs are most likely to become outliers. This pattern could be a result of using arithmetic mean aggregation.

There are a number of potential aspects of index construction that we have not fully explored in this analysis. For example, given the uniqueness of the specific data set and the rich spatial data, we did not explore the effect of spatial-temporal correlations in the different steps of the CI construction, i.e. data imputation.
Similarly, with the addition of new years' data, imputation methods could benefit from the longer time series. 
The spatial component could be potentially also exploited to construct a `spatial composite index'~\cite{trogu2018towards,siegel2016developing, fusco2018spatial,saib2015building}. Nevertheless, the experimental Health Index fulfills the criteria advocated by Ashraf et al. \cite{ashraf2019population} and it constitutes a  starting basis for statistical improvements that will improve the feature releases. 

\subsection*{Aftermath} 

In summary, this analysis conducted on the 2015-18 beta Health Index  served as proof on the statistical coherence and investigated choices and issues that arise in the  process of building a new composite index. 
The aim of the ONS Health Index is to become a reliable harmonized index over time and over space, with inclusion  of finer geographical level and potential stratification of population by age and sex, including Scotland and Wales. The suggestions highlighted in this article are not exhaustive  due to the current evolving nature of the index, but provide a valuable tool that  serves as guidance for the upcoming versions.

While we assessed the beta version, the Health Index version from year 2019  already includes the Lower Tier Local Authorities (LTLA)  for England (307 LTLAs) and several suggestions have been integrated. We review the propositions made in this paper and how these have been included.

\begin{itemize}
    \item  Data is winsorized to derive weights by factor analysis, reducing need for much other transformation to normalize. The final data,  which gets aggregated later on to produce Index scores,  is not winsorized so LTLAs with extreme values can still observe change over time. 

    \item  Public green space indicator - that was negatively correlated with private outdoor space - will be removed in the 2020 Health Index.
   
    \item The passage to LTLA relaxed the correlation seen  for Behavioural risk factors, and drug misuse indicators has been changed to a different source to meet the granularity criteria.
    
    \item Subdomains `Risk factors for children' and `Children and young people's education' are now joined into a unique subdomain `Children and young people'.
    
    \item Hypertension (renamed to `High blood pressure') continues to show high positive correlation with  cardiovascular, respiratory and musculoskeletal conditions, even at LTLA level. Currently, the ONS is exploring potential alternatives for these indicators.

    \item At LTLA level data no longer present a negative correlation between Healthy Places and the other domains: they are now all positively correlated, albeit Places has a weaker correlation with the others than the Lives--People correlation. 
    
    \item The inclusion of a smaller geography as base unit and changes observed for in correlations pairs, changes the derived weights that account for the correlation.  The added value of our analysis highlighted  how crucial the weights are, not only in terms of value but also in terms of interpretation. The ONS is going to evaluate the possibility to have non-compensatory weights defined by expert opinion, to better capture what stakeholders rate as important among the domains and subdomains. 
    
    \item We argued that between FA and PCA derived weights, PCA should be preferred. The ONS has not implemented this  proposed modification. This is due to the evolving nature of the index at this stage. The extension to  additional geographical layer and population stratification will have an impact on the correlation matrices and therefore  weights may undergo substantial changes, before a stable version is finalized. 
    
    \item  Finally, our last suggestion promoted a geometric mean aggregation formula, that implies only partial substitution and rewards balance by penalizing uneven performance in the underlying domains.  Initially, among the principle followed by the ONS in producing a composite index, there was the necessity to be able to have simple statistical methods that could be easy to explain. However, in the course of this analysis, we showed that the geometric mean offers a valid alternative, used by many other well-known and respected indexes such as the Human Development Index~\cite{hdi}. This option will be taken into account in the ONS's upcoming methodological evaluations. 
    
\end{itemize}

\subsection*{Conclusion}

In conclusion, the ONS Health Index (2015--18) presents a summary of the health of the population of England and fills a gap in policy making and assessment tools. 
The index is based on a hierarchical geographical structure, starting from the Upper Tier Local Authority level, rising to National level. It provides a detailed and flexible composite measurement, that will allow policy makers to assess changes in population health, and to plan interventions by identifying areas and policy domains where interventions can provide significant, quantifiable impact. Future Health Index editions, with finer geographic granularity and population subgroups, will enrich the understanding of  health determinants and guide bespoke interventions and assessments.

\newpage

\section*{Acknowledgments}

AFS is grateful to William Becker for his help and for writing the R-package COINr and to Professor Avi Feller for useful comments.

\bibliographystyle{alpha}
\bibliography{compsiteb}

\end{document}


\maketitle

\section{The ONS Experimental Health Index}
\label{sec:ONS}

\begin{figure}[ht]
\centering
\includegraphics[width=0.95\textwidth]{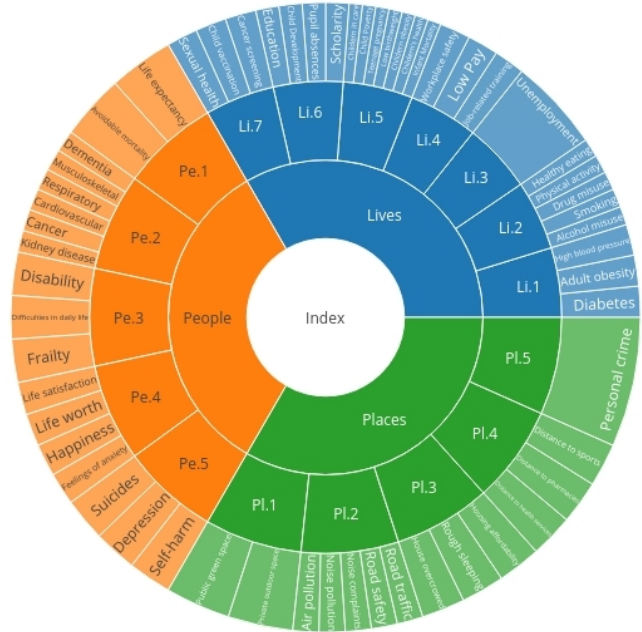}
\caption{\label{fig:fig1}The Health Index structure.}
\end{figure}

The annual Office of National Statistics (ONS)  Experimental Health index(HI) is built on data collected in the 149 Upper Tier Local Authorities (UTLAs) in England (United Kingdom), from 2015 to 2018. A UTLA is composed by counties, metropolitan counties, inner and outer London, unitary authorities, with a 2015 population range from  38,582 to 1,523,100.
The Index is composed by three main domains: Healthy people, Healthy places, Healthy lives (see Figure \ref{fig:fig1}). Each domain represents a dimension, and they are further split in 17 subdomains, defined over 58 indicators.
The theoretical framework that led to the choice of the indicators and the definitions of the subdomains and domains have been discussed elsewhere \cite{ONS}. The main outcome is the overall  health index currently  computed for England only ( in the future it will include Scotland and Wales).  Given the index structure,  the  health index  is  also available at subdomains and domains levels, at UTLA, regional and national level. The index structure composed by indicators, subdomains and domains, is show in figure \ref{fig:fig1} and  full details in table \ref{tab:SM1.1}.

\clearpage
\begin{landscape}
\centering
\begin{longtable}{llll}
\toprule
        Domains & Subdomains & Indicators & Labels \\ \hline
        Lives & Li.1  & Prevalence of diabetes & diab \\ 
        ~ & Physiological risk factors & Percentage of adults (aged 18+) classified as overweight or obese modelled and age-standardised & ow\_obs \\
        ~ & & Prevalence of hypertension & hyp\_ten \\ \hline
        ~ & Li.2  & Admission episodes for alcohol-related conditions (Narrow) & alc \\ 
        ~ & Behavioural risk factors & Smoking Prevalence in adults (18+) - current smokers & smoke \\ 
        ~ &  & Drug misuse & drug \\ 
        ~ &  & Percentage of physically active adults ($>$150 minutes/week) & activity \\ 
        ~ &  & Proportion of the population meeting the recommended '5-a-day' on a 'usual day' (adults) & eating \\ \hline
        ~ & Li.3 Unemployment & Model-based estimates of unemployment & unempl \\ \hline
        ~ & Li.4 & Percentage of working age adults (16 to 64) who received job-related training in the last 13 weeks & training \\ 
        ~ & Working conditions & \% of jobs earning below NLW & low\_pay \\
        ~ &  & RIDDOR reported non-fatal injuries & wp\_safety \\ \hline
        ~ & Li.5 & Infant mortality & inf\_mort \\ 
        ~ & Risk factors for children & \% of school pupils with social, emotional and mental health needs & semh\_child \\
        ~ &  & Overweight and obesity & ow\_obs\_child \\ 
        ~ &  & Low birth weight of term babies (less than 2500g) & low\_bw \\ 
        ~ & & Conceptions in women aged under 18 per 1,000 females aged 15-17 & teen\_preg \\ 
        ~ &  & Children in absolute low income & pov\_child \\ 
        ~ &  & Rate of children in state care & incare\_child \\ \hline
        ~ & Li.6 & Proportion of pupils in sustained education, employment or training in the year after KS4 & in\_eet \\ 
        ~ & Children and young people’s education & Percentage of persistent absentees & pupil\_abs \\ 
        ~ &  & Percentage of students (5 years old) achieving a good level of development & ey\_dev \\ 
        ~ & & Percentage of pupils achieving grades 4 or above in English and Mathematics GCSEs KS4 & ks4\_dev \\ \hline
        ~ & Li.7 & Cancer screening & can\_screen \\ 
        ~ & Protective measures & Population vaccination coverage - MMR for two doses (5 years old) & vac \\ 
        ~ &  & New STI diagnoses (exc chlamydia aged $>$ 25) / 100,000 & sti\_screen \\ \hline
        People & Pe.1 & Life expectancy at birth & hle \\ 
        ~ & Mortality & Avoidable deaths & avoid\_mort \\ \hline
        ~ & Pe.2 & Prevalence of dementia & dementia \\ 
        ~ & Physical health conditions & Prevalence of musculoskeletal disease & musculo \\ 
        ~ & & Prevalence of respiratory disease & resp \\ 
        ~ &  & Prevalence of cardiovascular disease & cvd \\ 
        ~ & & Prevalence of cancer & cancer \\ 
        ~ &  & Prevalence of chronic kidney disease & kidney \\ \hline
        ~ & Pe.3 & Disability & disability \\ 
        ~ & Difficulties in daily life& Percentage with long-term condition which reduces ability to carry out day-to-day activities & diff\_adls \\ 
        ~ & & Hip fractures in people & frailty \\ \hline
        ~ & Pe.4 & Mean satisfaction score & srwb\_sat \\ 
        ~ & Personal well-being & Mean worthwhile score & srwb\_worth \\ 
        ~ &  & Mean happiness score & srwb\_hap \\ 
        ~ &  & Mean anxiety score & srwb\_anx \\ \hline
        ~ & Pe.5 & Suicides & suicide \\ 
        ~ & Mental health & Depression & depres \\ 
        ~ &  & Hospital admissions as a result of self harm & self\_harm \\ \hline
        Places & Pl.1 & Average distance to nearest Park, Public Garden & pub\_greenspace \\ 
        ~ & Access to green space & Access to garden space & priv\_greenspace \\ \hline
        ~ & Pl.2 & Air pollution & air\_poll \\ 
        ~ & Local environment & Noise pollution & trans\_noise \\ 
        ~ &  & Rate of complaints about noise & env\_noise \\ 
        ~ & & Road Safety & road\_safety \\ 
        ~ &  & Road traffic & road\_traff \\ \hline
        ~ & Pl.3 & Household overcrowding & house\_overcrowd \\ 
        ~ & Access to housing& Number of people sleeping rough & rough\_sleep \\ 
        ~ &  & Housing affordability & house\_afford \\ \hline
        ~ & Pl.4 & Distance/travel time to nearest General Practice & GP\_dist \\ 
        ~ & Access to services& Distance/travel time to nearest pharmacy (dispensary) & pharm\_dist \\ 
        ~ &  & Distance/travel time to nearest sports or leisure facility & sport\_dist \\ \hline
        ~ & Pl.5 Crime & Police recorded personal crime & crime\_pers \\ \hline
\caption{Health Index structure and descriptions.}
\label{tab:SM1.1}
\end{longtable}
\end{landscape}
\clearpage

\section{Health Index Construction}
\label{sec:ci_steps}

The health index is built starting from a tensor $\mathcal X$ of raw data, with elements $x_{cit}$.
Here,
each $c \in C$ is an upper tier local authority (UTLA), for the set $C$ of $|C|=149$ UTLAs;
each $i \in I$ is an indicator, for the set $I$ of $|I|=58$ indicators;
and each $t \in T = \{2015, 2016, 2017, 2018\}$ denotes the year.
We are also given a partition of the set of UTLAs, $C$, into a set $R$ of $|R|=9$ regions, $r \in R$, which are disjoint subsets $r \subseteq C$ of UTLAs.

Here, we review the composite index steps taken to built the HI by Office for National Statistics:
\begin{itemize}
    \item Imputation
    \item Data treatment
    \item Normalization
    \item Weighted aggregation into subdomains, domains, overall index value
    \item Weighted geographical aggregation into regions and the nation
\end{itemize}

We describe in details each of these steps.

\subsection{Imputation}
\label{sec:imputation}
The first stage of HI construction is to impute data that is missing from the tensor, $\mathcal X = (x_{cit})$.
The imputation process is carried out as follows:
\begin{enumerate}
 \item  If $x_{cit}$ is missing but $x_{cit_-}$ and $x_{cit_+}$ exist for $t_- < t < t_+$, such that we have data either side of a missing value, then the missing value, $x_{cit}$, is calculated as a linear interpolation of $x_{cit_-}$ and $x_{cit_+}$.
  \item  If $x_{cit}$ is missing and only exactly one of $x_{cit_-}$ or $x_{cit_+}$ exists, then the missing value, $x_{cit}$, is calculated as the nearest value in the time series, whichever of  $x_{cit_-}$ or $x_{cit_+}$ exists.
  \item  If $x_{cit}$ is missing for all $t$, they impute the mean across that UTLA's corresponding region, $r \ni c$. That is, we calculate $x_{cit} = \sum_{c' \in r \setminus \{c\}} x_{c'it} / (|r|-1)$. This only occurred for two UTLAs, for one indicator.
  \item If $x_{cit}$ is suppressed because the numerator is small, that is, the value is too low to be presented, then it was replaced with the lowest value presented for that data series.
  \item If $x_{cit}$ is suppressed because the denominator is small, that is, there are too few observations to base the data on, then it was replaced with the median value from the data series.
\end{enumerate}
As shown in Table \ref{tab:tab1}, ten indicators were missing across all UTLAs (i.e. for all $c \in C$) for certain years. For example, $\texttt{pub-greenspace}$ (representing \emph{access to public green space}) was available only for 2018, and $\texttt{trans-noise}$ noise pollution, only for 2016.
In this case, both are imputed to be constant over the four year span.

\begin{table}
  \centering
    \begin{threeparttable}[b]
  \caption{\label{tab:tab1}Indicators completely missed by year.}
\begin{tabular}{llcccc}
\toprule 
Indicator $i$ & Description & 2015 & 2016 & 2017 & 2018 \\ 
\midrule
 \texttt{diff-adls} & Difficulty in daily activities & \xmark & \xmark & \xmark & \\
 \texttt{ow-obs} & Obesity\tnote{1} & \xmark & & & \\
 \texttt{activity} & Physical activity\tnote{2} & \xmark & & & \\
 \texttt{eating} & Eating\tnote{3}  & \xmark & & & \\
 \texttt{in-eet} & Scholarity\tnote{4} & \xmark & & & \\
 \texttt{pub-greenspace} & Public green space  & \xmark & \xmark & \xmark & \\
 \texttt{priv-greenspace} & Private green space  & \xmark & \xmark & \xmark & \\
 \texttt{house-overcrowd} & Household overcrowding  & & \xmark & \xmark & \xmark \\
 \texttt{trans-noise} & Noise pollution & \xmark & & \xmark & \xmark \\
 \texttt{env-noise} & Noise complaints& & \xmark & \xmark & \\
\bottomrule
\end{tabular}
 \begin{tablenotes}
    \item[1]Percentage of adults (aged $18+$) classified as overweight or obese
    \item[2]Percentage of physically active adults ($>150$ min activity per week)
    \item[3]Proportion of the adult population meeting the recommended `5-a-day' on a `usual day'
    \item[4]Proportion of pupils in sustained education, employment or training in the year after KS4
  \end{tablenotes}
\end{threeparttable}
\end{table}

\subsection{Data Treatment}
\label{sec:treatment}

Once the missing data has been imputed, the tensor $\mathcal X = (x_{cit})$ is decomposed into $|I|=58$ flattened data sets (all annual data is collapsed in one matrix) , $\mm{X}_i = \{ x_{cit} : c \in C,t \in T \}$ for each $i \in I$.

The skewness and kurtosis of each $\mm{X}_i$ is computed.
Using the data transformations $f_i$ listed in Table \ref{tab:SM1.5}, data is transformed  $\mm X_i$ to $\mm Y_i = \{ y_{cit} = f_i(x_{cit}) : c \in C,t \in T \}$.

The assignment of each indicator, $i$, to a chosen transformation, $f_i$, is selected to minimise the absolute values of skewness and kurtosis of $\mm Y_i$, aiming for absolute skewness $\leq 2$ and absolute kurtosis $\leq 3.5$. 
By minimising (absolute) skewness and kurtosis, we aim to ensure that the transformed data $\mm Y_i$ is approximately normally distributed.
Note that, for the 18 indicators $i$ \emph{not} listed in Table \ref{tab:SM1.5}, the skewness and kurtosis of $\mm X_i$ is optimal, compared to all the possible transformations considered.

\afterpage{
\clearpage
\begin{landscape}
\centering
\begin{longtable}{lcccccc}
\toprule
Indicator $i$ & \multicolumn{2}{c}{Median (IQR)} & \multicolumn{2}{c}{Skewness} & \multicolumn{2}{c}{Kurtosis} \\
\cmidrule(lr){2-3} \cmidrule(lr){4-5} \cmidrule(lr){6-7}
& $\mm X_i$ & $\mm Y_i$ & $\mm X_i$ & $\mm Y_i$ & $\mm X_i$ & $\mm Y_i$ \\
\midrule \multicolumn{7}{l}{\bf Data transform: $x \mapsto \log(x)$} \\
\texttt{avoid-mort} & 240 (204, 278) & 5.48 (5.32, 5.63) & 1 & 0.10 & 0 & -0.59 \\
\texttt{suicide} & 9.84 (8.50, 11.20) & 2.29 (2.14, 2.42) & 0.73 & 0.02 & 1.03 & 0.15 \\
\texttt{musculo} & 0.60 (0.50, 0.72) & -0.52 (-0.69, -0.32) & 1.00 & 0.03 & 1.87 & 0.05 \\
\texttt{srwb-anx} & 2.90 (2.76, 3.06) & 1.06 (1.02, 1.12) & 0.49 & 0.23 & 0.42 & 0.17 \\ 
\texttt{Alcohol consumption} & 641 (558, 753) & 6.46 (6.32, 6.62) & 1 & 0.15 & 0 & -0.46 \\
\texttt{drug} & 40 (29, 55) & 3.69 (3.37, 4.01) & 1 & 0.09 & 3 & -0.09 \\ 
\texttt{Unemployment} & 4.80 (3.80, 5.90) & 1.57 (1.34, 1.77) & 0.74 & 0.00 & 0.51 & -0.48 \\
\texttt{incare-child} & 65 (46, 82) & 4.18 (3.83, 4.41) & 1 & -0.14 & 1 & -0.42 \\ 
\texttt{Infant mortality} & 3.63 (3.10, 4.46) & 1.29 (1.13, 1.49) & 0.95 & 0.02 & 1.37 & 0.22 \\ 
\texttt{Public greenspace} & 687 (466, 1,053) & 6.53 (6.14, 6.96) & 1 & 0.24 & 2 & -0.62 \\ 
\texttt{Noise pollution} & 5 (4, 7) & 1.56 (1.30, 2.01) & 4 & 1.10 & 22 & 1.81 \\ 
\texttt{Noise complaints} & 6.2 (4.4, 9.2) & 1.83 (1.48, 2.22)& 1.5 & -0.02 & 2.6 & -0.02 \\ 
\texttt{road-safety} & 405 (290, 730) & 6.00 (5.67, 6.59) & 2 & 0.74 & 6 & 0.12 \\ 
\texttt{rough-sleep} & 5.6 (2.9, 10.0) & 1.71 (1.05, 2.31) & 6.0 & -0.24 & 48.5 & 0.38 \\ 
\texttt{House affordability} & 7.9 (5.6, 11.1) & 2.07 (1.72, 2.41) & 1.4 & 0.33 & 2.5 & -0.63 \\ 
\texttt{Distance to GP} & 0.68 (0.54, 0.97) & -0.39 (-0.62, -0.03) & 3.35 & 0.73 & 16.82 & 1.08 \\ 
\texttt{Distance to Pharmacy} & 0.63 (0.52, 0.88) & -0.46 (-0.65, -0.12) & 3.80 & 0.92 & 19.65 & 1.80 \\ 
\texttt{sport-dist} & 0.44 (0.40, 0.51) & -0.82 (-0.92, -0.67) & 0.91 & -0.12 & 2.04 & 1.49 \\ 
\midrule \multicolumn{7}{l}{\bf Data transform: $x \mapsto x^{1/2}$} \\
\texttt{depres} & 9.34 (7.95, 10.93) & 3.06 (2.82, 3.31) & 0.35 & 0.01 & 0.12 & -0.12 \\ 
\texttt{self-harm} & 195 (142, 245) & 14.0 (11.9, 15.7) & 1 & -0.1 & 1 & -0.2 \\ 
\texttt{Teen pregnancy} & 19 (14, 24) & 4.35 (3.79, 4.90) & 0 & -0.04 & 0 & -0.08 \\ 
\texttt{in-eet} & 2.80 (2.00, 3.70) & 1.67 (1.41, 1.92) & 0.58 & 0.03 & -0.01 & -0.12 \\ 
\texttt{semh-child} & 14.90 (13.68, 16.52) & 3.86 (3.70, 4.06) & 0.20 & -0.08 0.54 & 0.45 \\ 
\texttt{road-traff} & 0.007 (0.003, 0.013) & 0.08 (0.05, 0.11) & 0.843 & 0.12 & 0.476 & -0.89 \\
\midrule \multicolumn{7}{l}{\bf Data transform: $x \mapsto x^{1/3}$} \\
\texttt{low-bw} & 2.75 (2.37, 3.20) & 1.40 (1.33, 1.47) & 0.61 & 0.10 & 0.63 & 0.30 \\ 
\texttt{Children poverty} & 14.6 (11.4, 18.5) & 2.44 (2.25, 2.65) & 0.7 & 0.12 & 0.1 & -0.33 \\ 
\texttt{wp-safety} & 283 (228, 335) & 6.57 (6.11, 6.95) & 1 & -0.01 & 2 & 0.82 \\ 
\midrule \multicolumn{7}{l}{\bf Data transform: $x \mapsto x^2$} \\
\texttt{Respiratory disease}
 & 4.07 (3.40, 4.44) & 16.5 (11.5, 19.7)& -0.45 & -0.1 & -0.67 & -0.7 \\ 
\texttt{Obesity children} & 28.7 (26.4, 30.7) & 824 (696, 941) & -0.2 & 0 & -0.3 & 0 \\ 
\texttt{Hypertension} & 14 (12.21, 15.67) & 196 (149, 246) & -0.44 & 0 & -0.45 & -1 \\ 
\midrule \multicolumn{7}{l}{\bf Data transform: $x \mapsto x^3$} \\
\texttt{Mean satisfaction score} & 7.65 (7.55, 7.76) & 448 (430, 467) & -0.42 & 0 & 0.07 & 0 \\ 
\texttt{Mean worthwhile score} & 7.85 (7.75, 7.93) & 484 (465, 499) & -0.55 & 0 & 0.24 & 0 \\ 
\texttt{Mean happiness score} & 7.51 (7.39, 7.60) & 424 (403, 439) & -0.40 & 0 & 0.21 & 0 \\ 
\texttt{Obesity} & 62 (59, 66) & 240,816 (201,907, 282,317)& -1 & 0 & 0 & 0 \\ 
\texttt{Cancer screening} & 70.6 (66.4, 72.5) & 351,168 (292,241, 380,487) & -0.9 & -1 & 0.7 & 0 \\ 
\texttt{Vaccination coverage} & 90.9 (87.1, 92.8) & 751,927 (660,592, 798,447) & -1.4 & -1 & 2.5 & 1 \\ 
\texttt{Good development \footnote{Percentage of students achieving a good level of development}} & 69.7 (66.3, 72.2) & 338,424 (291,556, 376,175) & -0.5 & 0 & 0.4 & 0 \\ 
\texttt{Private greenspace} & 90 (85.9, 91.7) & 730,193 (633,840, 771,095) & -2.2 & -2 & 5.8 & 3 \\ 
\midrule \multicolumn{7}{l}{\bf Data transform: $x \mapsto -1/x$} \\
\texttt{New STI\footnote{Sexually Transmitted Infections}diagnoses}& 682 (556, 916) & -0.0015 (-0.0018, -0.0011) & 3 & 0.1430 & 7 & -0.0817 \\ 
\texttt{Percentage of house-overcrowd} & 7 (5, 11) & -0.15 (-0.21, -0.09) & 2 & -0.25 & 2 & -0.48 \\ 
\bottomrule
\captionsetup{width=\columnwidth}
\caption{Transformation of 40 raw indicator values $x_{cit}$ into $y_{cit}$. The 18 remaining indicators not shown in this list are not transformed, such that $y_{cit} = x_{cit}$.}
\label{tab:SM1.5}
\end{longtable}
\end{landscape}
\clearpage
}

In Table \ref{tab:SM1.5}, we observe that, of the 40 transformed indicators, the skewness and kurtosis of $\mm X_i$ were already within the desired limits (see, for example \texttt{avoid-mort}, \texttt{ow-ob-child} or \texttt{incare-child}).
Furthermore, we observe in Table \ref{tab:SM1.5} that 18 indicators have been log transformed.
As reported by \cite{nardo2005tools}, ``\emph{when the weighted variables in a linear aggregation are expressed in logarithms, this is equivalent to the geometric aggregation of the variables without logarithms. The ratio between two weights indicates the percentage improvement in one indicator that would compensate for a one percentage point decline in another indicator. This transformation leads to attributing higher weight for a one unit improvement starting from a low level of performance, compared to an identical improvement starting from a high level of performance.}''

\subsection{Normalization}
\label{sec:nor}

In the preceding step, for each indicator, $i$, we have determined a data transform function, $f_i$, chosen to ensure normality of the observed data.
Using these functions, we can transform the raw data tensor $\mathcal X$ into a new data tensor, $\mathcal Y$, with elements $y_{cit} = f_i(x_{cit})$.
In this step, we further transform the data to ensure that each indicator is directly comparable.
This is done by standardising each indicator, to produce $z$ scores.
At this stage, we also address the polarity of the $z$ scores, multiplying by $-1$ as necessary, in order to ensure that larger positive numbers always correspond to healthier outcomes in each of the indicators. 

The normalization step in the ONS Health Index accounts for time and geography, and allows indicators to be compared on the same scale, weighting by the UTLA populations.
The normalization transforms elements $y_{cit}$ of $\mathcal Y$ into z-scores, 
\[
z_{cit} = (-1)^{\delta_i} \left[ \frac{y_{cit} - \mu_i}{\sigma_i} \right],
\]
which then form the elements of the tensor $\mathcal Z = (z_{cit})$.
For each indicator, $i$, we specify $\delta_i = 0$ or $\delta_i = 1$ to ensure that larger positive values for $z_{cit}$ correspond to improved health, a property which we term as being \emph{health directed}. 
The translation and scaling parameters, $\mu_i$ and $\sigma_i$, respectively, are constructed based solely on the transformed data from the baseline year $t=2015$ as 
\begin{align*}
    \mu_i &= \sum_{\substack{t=2015 \\ c \in C}} w_{ct} y_{cit}, \\
    \sigma_i^2 &= \frac{C}{C-1} \sum_{\substack{t=2015 \\ c \in C}} w_{ct} (y_{cit} - \mu_i)^2,
\end{align*}
where each normalised UTLA weight, $w_{ct}$, for $c \in C$ and $t \in T$, is the percentage of the national population in UTLA $c$ in year $t$, so that $\sum_{c \in C} w_{ct} = 1$ for each $t$, by definition.

Finally, given the $z$-scores $z_{cit}$ forming the tensor $\mathcal Z$, the ONS Health Index presents the $z$-scores as Health Index values, 
\[
h_{cit} = H(z_{cit}) = 100 + 10 z_{cit},
\]
which are translated and rescaled $z$-scores, such that $h_{cit} = 100$ means that the transformed value, $y_{cit}$, for indicator $i$ in the UTLA $c$ in year $t$ is equal to the weighted mean, $\mu_i$. 
Similarly, increments of $\pm 10$ on $h_{cit}$ correspond to increments of $\pm (-1)^{\delta_i} \sigma_i$ on $y_{cit}$, or one health-directed standard deviation.
As with the tensors $\mathcal X, \mathcal Y, \mathcal Z$, we can denote the tensor of Health Index values by $\mathcal H = (h_{cit})$.

\subsection{Indicator Aggregation method}
The characteristic feature of the ONS Health Index is that a hierarchical structure is imposed on the set of indicators, $I$.
Each indicator $i \in I$ belongs to exactly one subdomain, $s \subseteq I$, which are collected in the set $s \in S$ of size $|S|=17$.
Similarly, each subdomain $s \in S$ belongs to exactly one domain, $d \subseteq S$, which are collected in the set $d \in D$ of size $|D|=3$.

Consider the tensor $\mathcal Z = (z_{cit})$ of health directed $z$-scores at the level of UTLA $c \in C$, indicator $i \in I$, year $t \in T$, constructed by the process outlined in Sections \ref{sec:imputation} to \ref{sec:nor}.
We will make this level explicit using subscript notation, writing $\mathcal Z_{CIT} = (z_{cit})$ for the tensor of $z$-scores at the UTLA--indicator level.
The ONS Health Index uses linear aggregation to aggregate the $z$-scores at subdomain and domain level, where
\begin{align*}
    z_{cst} &= \alpha_s \sum_{i \in s} w_i z_{cit}, \\
    z_{cdt} &= \alpha_d \sum_{s \in d} w_s z_{cst},
\end{align*}
where $w_i$ and $w_s$ are indicator and subdomain weightings, respectively,
and where $\alpha_s$ and  $\alpha_d$ are normalisation constants.
These $z$-scores form the tensors $\mathcal Z_{CST} = (z_{cst})$ and $\mathcal Z_{CDT} = (z_{cdt})$ at the UTLA--subdomain and UTLA--domain level, respectively.
Finally, an overall health index $z$-score can be formed by a final aggregation into
\[
z_{ct} = \alpha \sum_{d \in D} w_d z_{cdt},
\]
for domain weightings $w_d$ and normalisation constant $\alpha$, giving a matrix $\mathcal Z_{CT} = (z_{ct})$ for the overall health index score, at the UTLA level.

Given the linear aggregation methodology from indicator to subdomain, to domain, to overall health index, there are two types of parameters to specify.
We need to choose the weightings, $w_i$, $w_s$, and $w_d$, of the indicators, subdomains and domains.
We also need to specify the normalisation constants, $\alpha_s$, $\alpha_d$, and $\alpha$.

Suppose that the weightings $w_i$, $w_s$, and $w_d$ have been specified.
Given these weightings, the ONS Health Index fixes the normalisation constants as
\begin{align*}
    \alpha_s^{-1} &= \sum_{i \in s} w_i, \\
    \alpha_d^{-1} &= \sum_{s \in d} w_s, \\
    \alpha^{-1} &= \sum_{d \in D} w_d.
\end{align*}
These choices of normalisation constant ensures that the aggregated $z$-scores at each of the subdomain, domain, and overall levels are simply weighted averages of the $z$-scores at the indicator, subdomain, and domain levels, respectively.

\subsection{Weighting}

Within this framework, independently by the of the choice of normalisation constant in the section above, we need to specify weights $w_i$, $w_s$, and $w_d$ for each of the indicators, subdomains and domains, uniquely up to a multiplicative constant.

\subsubsection{Subdomain and Domain weights}

In the ONS HI, there are two main sets of weights:  one for indicators and one for  domains and subdomains. At the  subdomain and domains levels, the choice is for equal weights. In detail for $w_s = 1$ for all $s \in S$, and $w_d = 1$ for all $d \in D$ (recalling that these weights are---currently---subsequently normalised by $\alpha_d = 1 / \sum_{s \in d} w_s = 1 / |d|$ and $\alpha = 1 / \sum_{d \in D} w_d = 1/3$, respectively).

\subsubsection{Indicator weights}

The weights $w_i$ for indicators $i \in I$ are chosen by applying factor analysis to each subdomain, $d \in D$, in turn.

Data after being normalized is then rescaled to a z-score (0,1) and all the 4 years are collapsed.  

For each  $d \in D$, we carry out a factor analysis on the indicators $i \in d$, 

$Z_{d, c \times t}= \sum a_{d} F_{d} +e_d$

where $e_d$.

The weights are chosen as the first loading factor, taken in absolute value. 
For example, for a subdomain $d = \{ i_1, i_2 \}$ comprised of two indicators, with first  factor loadings are 0.5 and 0.7,  then set the weights $w_{i_1} = 0.5$ and $w_{i_2} = 0.7$, recalling that these weights are normalised, through multiplying by $\alpha_d = 1/1.2$.

\subsection{Geographical Aggregation}

Aggregation can also take place in the geographical dimension.
We recall that each UTLA, $c \in C$, is a member of exactly one region, $r \subseteq C$, which are collected into the set $R$ of $|R|=9$ regions, $r \in R$.
We further recall that the weights $w_{ct}$ for $c \in C$ and $t \in T$ correspond to the percentage of the national population in UTLA $c$ in year $t$.
Thus, we can aggregate to the regional level,
\begin{align*}
    z_{rit} &= \sum_{c \in r} w_{ct} z_{cit}, \\
    z_{rst} &= \sum_{c \in r} w_{ct} z_{cst} = \alpha_s \sum_{i \in s} w_i z_{rit}, \\
    z_{rdt} &= \sum_{c \in r} w_{ct} z_{cdt} = \alpha_d \sum_{s \in d} w_s z_{rst}, \\
    z_{rt} &= \sum_{c \in r} w_{ct} z_{ct} = \alpha \sum_{d \in D} w_d z_{rdt},
\end{align*}
for each of the indicator, subdomain, domain, and overall level $z$-scores.
Thus, we can form tensors $\mathcal Z_{RIT} = (z_{rit})$, $\mathcal Z_{RST} = (z_{rst})$ and $\mathcal Z_{RDT} = (z_{rdt})$ at the regional--indicator, regional--subdomain, and regional--domain level, respectively, and the matrix $\mathcal Z_{RT} = (z_{rt})$ at the regional--overall level.

Finally, we can also aggregate to a national level, writing
\begin{align*}
    z_{it} &= \sum_{c \in C} w_{ct} z_{cit}, \\
    z_{st} &= \sum_{c \in C} w_{ct} z_{cst} = \alpha_s \sum_{i \in s} w_i z_{it}, \\
    z_{dt} &= \sum_{c \in C} w_{ct} z_{cdt} = \alpha_d \sum_{s \in d} w_s z_{st}, \\
    z_{t} &= \sum_{c \in C} w_{ct} z_{ct} = \alpha \sum_{d \in D} w_d z_{dt},
\end{align*}
for each of the indicator, subdomain, domain, and overall level $z$-scores.
Thus, we can form matrices $\mathcal Z_{IT} = (z_{it})$, $\mathcal Z_{ST} = (z_{st})$ and $\mathcal Z_{DT} = (z_{dt})$ at the national--indicator, national--subdomain, and national--domain level, respectively, and the vector $\mathcal Z_{T} = (z_{t})$ at the national--overall level.

A key observation from the definitions above is that, due to the linearity of both geographical aggregation and indicator aggregation, the aggregation order is interchangeable.
For example, the regional--subdomain level aggregation, $\mathcal Z_{RST}$, can be achieved
\begin{itemize}
\item by aggregating from UTLA--indicator level to UTLA--subdomain level and then to regional--subdomain level;
\item or by aggregating from UTLA--indicator level to regional--indicator level and then to regional--subdomain level;
\end{itemize}
achieving the same results.

\section{A modified Health Index version}
\label{sec:smMod}

For the modified version we carried out the following steps:
\begin{itemize}
    \item Treatment: winsorization was applied as data treatment, we  have minimized the number of indicators that where transformed to less than 7 per year, see table \ref{tab:SM.data}.
    \item  Weights and aggregation: arithmetic mean and equal weights. 
\end{itemize}

We have evaluated the data by checking kurtosis and skewness, based on these two statistics, first we have  the distributions, by  reassigning outlying points to the next highest point. Then if still need a data transformation was carried out. In table \ref{tab:SM.data}, we have reported the variables treated and the all belong to Places, except for 'new STI diagnoses'. 

\begin{sidewaysfigure}[ht]
    \includegraphics[width=\textwidth]{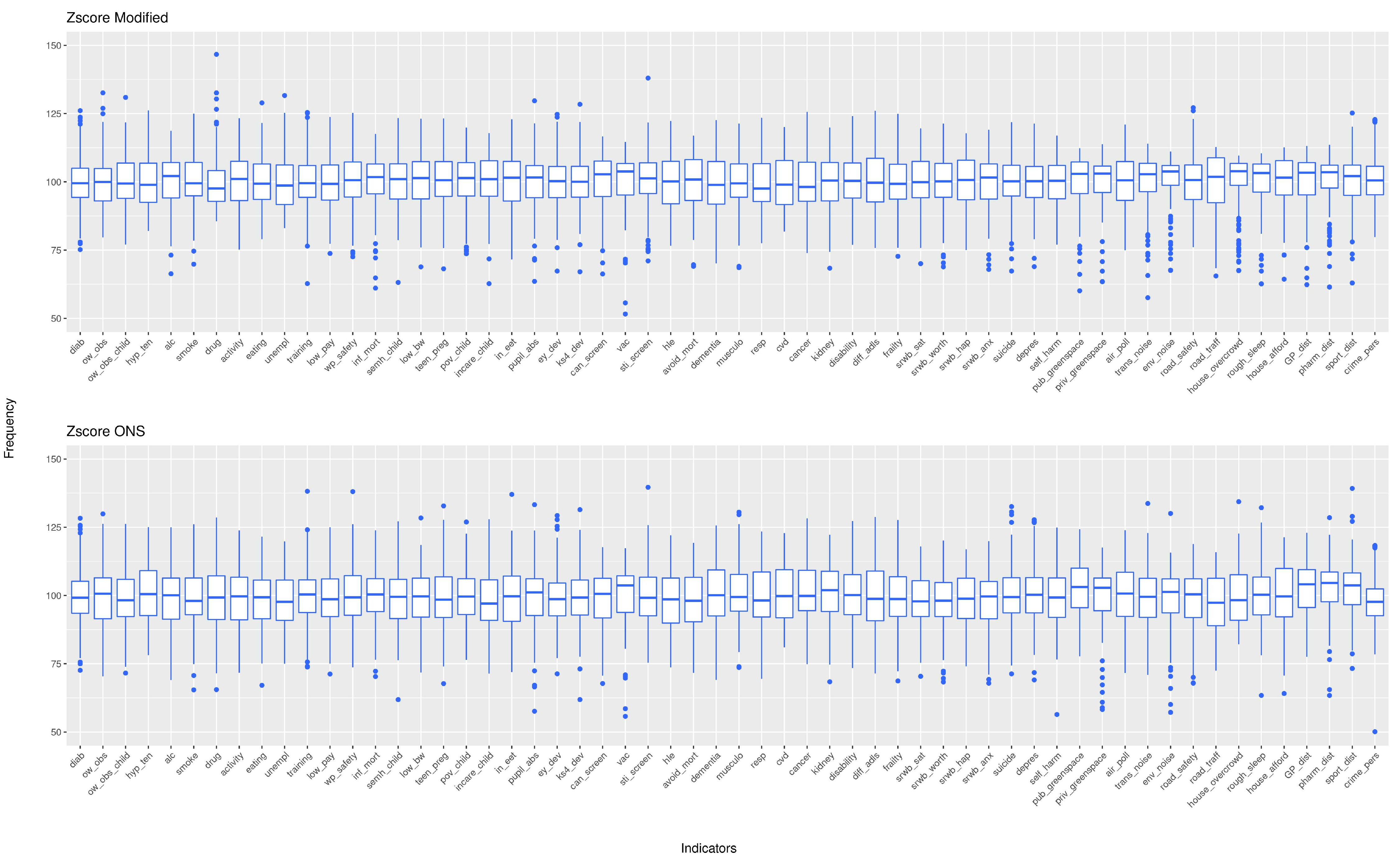}
    \caption{Indicators box plots (2015)  for ONS z-score and modified z-score transformations.}
    \label{fig:boxpl}
\end{sidewaysfigure}

We compared the boxplot distribution between the  the ONS z-score and the modified z-score ( see figure \ref{fig:boxpl}). Because we opted for winsorization, the modified distribution presents more indicators with outliers, in particular for values below the 25\% percentile compared to the ONS normalized indicators. 

We then compared the ranking, overall index values for the original ONS and the modified version, including the differences between these. 
 We notice that the values difference are within the range of $\pm 4$, but the rankings are shifting some UTLA of several positions, in particular the UTLA that are allocated in the middle, as tails remain stable. The shift in the ranks are due mostly to the ranking enforcement/constraints ( unique rank value, not shared ranking were allowed), but values wise, the shift are less dramatic, see table~\ref{tab:long1}.

\FloatBarrier
\begin{table}
  \centering
    \begin{threeparttable}
  \caption{\label{tab:SM.data}Data treatment by year.}
\begin{tabular}{llc}
\toprule 
Year & Indicator  & Data transform\\  
\midrule
    \multirow{5}{*}{2015}&New STI diagnoses& $x \mapsto \log(x)$  \\ 
 &Access to garden space & Winsorised 3 points  \\ 
 &Rate of complaints about noise & Winsorised 5 points \\ 
 &Road Safety & $x \mapsto \log(x)$\\ 
 &Number of people sleeping rough & Winsorised 2 points \\ 
 & Distance/travel time to nearest pharmacy (dispensary)& Winsorised 2 points  \\ \hline
  \multirow{7}{*}{2016}&New STI diagnoses & $x \mapsto \log(x)$  \\  
   & Access to garden space & Winsorised 3 points  \\ 
  & Rate of complaints about noise & $x \mapsto \log(x)$  \\ 
  & Road Safety & Winsorised 5 points \\ 
 & Number of people sleeping rough & Winsorised 4 points  \\ 
 & Distance/travel time to nearest GP practice & Winsorised 2 points  \\ 
 & Distance/travel time to nearest pharmacy (dispensary) & Winsorised 2 points \\ \hline
  \multirow{7}{*}{2017}& New STI diagnoses & Winsorised 3 points \\ 
 & Access to garden space & Winsorised 3 points \\ 
 & Rate of complaints about noise & $x \mapsto \log(x)$  \\ 
& Road Safety & $x \mapsto \log(x)$  \\ 
&Number of people sleeping rough &$x \mapsto \log(x)$   \\ 
& Distance/travel time to nearest GP practice & Winsorised 2 points \\ 
   & Distance/travel time to nearest pharmacy (dispensary) & Winsorised 2 points \\ \hline
\multirow{7}{*}{2018}& New STI diagnoses & Winsorised 4 points  \\ 
   & Access to garden space & Winsorised 3 points  \\ 
 & Rate of complaints about noise & $x \mapsto \log(x)$  \\ 
 & Road Safety &$x \mapsto \log(x)$\\ 
  & Number of people sleeping rough & Winsorised 2 points  \\ 
 & Distance/travel time to nearest GP practice & Winsorised 2 points  \\ 
  &Distance/travel time to nearest pharmacy (dispensary) & Winsorised 2 points  \\ 
 
\bottomrule
\end{tabular}
\end{threeparttable}
\end{table}

\begin{longtable}{lllllll}
\hline
        Unit Name & ONS  & Modified  & Rank  & ONS  &Modified  & Index  \\ 
        & Ranking &  Ranking & difference & index  &  index &  difference \\ 
        \hline
        Blackpool & 1 & 1 & 0 & 87.2 & 90.13 & 2.93 \\  
        Kingston upon Hull, City of & 2 & 2 & 0 & 91.78 & 95.31 & 3.52 \\  
        Wolverhampton & 3 & 12 & 9 & 92.4 & 96.93 & 4.53 \\  
        Liverpool & 4 & 6 & 2 & 92.84 & 95.96 & 3.12 \\  
        Knowsley & 5 & 3 & -2 & 93.1 & 95.56 & 2.46 \\  
        Manchester & 6 & 4 & -2 & 93.26 & 95.7 & 2.44 \\  
        Middlesbrough & 7 & 22 & 15 & 93.63 & 97.78 & 4.15 \\  
        Stoke-on-Trent & 8 & 7 & -1 & 93.99 & 96.19 & 2.2 \\  
        Nottingham & 9 & 14 & 5 & 94.02 & 97.19 & 3.17 \\  
        Salford & 10 & 8 & -2 & 94.26 & 96.49 & 2.23 \\  
        Rochdale & 11 & 10 & -1 & 94.38 & 96.74 & 2.36 \\  
        St. Helens & 12 & 9 & -3 & 94.38 & 96.68 & 2.3 \\  
        Hartlepool & 13 & 31 & 18 & 94.48 & 98.3 & 3.82 \\  
        Sandwell & 14 & 25 & 11 & 94.76 & 97.92 & 3.16 \\  
        Walsall & 15 & 28 & 13 & 95 & 98.14 & 3.13 \\  
        Tameside & 16 & 11 & -5 & 95.01 & 96.92 & 1.91 \\  
        Oldham & 17 & 24 & 7 & 95.04 & 97.79 & 2.75 \\  
        North East Lincolnshire & 18 & 21 & 3 & 95.35 & 97.72 & 2.37 \\  
        Doncaster & 19 & 27 & 8 & 95.46 & 97.98 & 2.52 \\  
        Newcastle upon Tyne & 20 & 37 & 17 & 95.5 & 98.44 & 2.94 \\  
        Bradford & 21 & 55 & 34 & 95.58 & 99.04 & 3.46 \\  
        Blackburn with Darwen & 22 & 44 & 22 & 95.74 & 98.69 & 2.95 \\  
        South Tyneside & 23 & 39 & 16 & 95.76 & 98.45 & 2.69 \\  
        Birmingham & 24 & 64 & 40 & 95.85 & 99.31 & 3.46 \\  
        Leicester & 25 & 26 & 1 & 95.87 & 97.96 & 2.09 \\  
        Rotherham & 26 & 16 & -10 & 95.88 & 97.35 & 1.46 \\  
        Sunderland & 27 & 36 & 9 & 95.92 & 98.42 & 2.5 \\  
        Redcar and Cleveland & 28 & 43 & 15 & 96.33 & 98.66 & 2.33 \\  
        Torbay & 29 & 13 & -16 & 96.36 & 97.13 & 0.77 \\  
        Southampton & 30 & 19 & -11 & 96.41 & 97.46 & 1.05 \\  
        Barnsley & 31 & 29 & -2 & 96.45 & 98.18 & 1.73 \\  
        Wakefield & 32 & 47 & 15 & 96.89 & 98.76 & 1.87 \\  
        Gateshead & 33 & 51 & 18 & 96.96 & 98.89 & 1.93 \\  
        County Durham & 34 & 20 & -14 & 97.01 & 97.7 & 0.7 \\  
        Barking and Dagenham & 35 & 84 & 49 & 97.08 & 100.01 & 2.93 \\  
        Halton & 36 & 48 & 12 & 97.11 & 98.8 & 1.69 \\  
        Sefton & 37 & 45 & 8 & 97.14 & 98.71 & 1.56 \\  
        Medway & 38 & 66 & 28 & 97.19 & 99.35 & 2.16 \\  
        Bolton & 39 & 63 & 24 & 97.19 & 99.29 & 2.09 \\  
        Derby & 40 & 41 & 1 & 97.19 & 98.55 & 1.36 \\  
        Bristol, City of & 41 & 18 & -23 & 97.2 & 97.42 & 0.22 \\  
        Stockton-on-Tees & 42 & 56 & 14 & 97.34 & 99.05 & 1.71 \\  
        Hackney & 43 & 15 & -28 & 97.34 & 97.23 & -0.11 \\  
        Dudley & 44 & 75 & 31 & 97.46 & 99.73 & 2.27 \\  
        Plymouth & 45 & 49 & 4 & 97.5 & 98.8 & 1.3 \\  
        Greenwich & 46 & 52 & 6 & 97.74 & 98.89 & 1.15 \\  
        Islington & 47 & 17 & -30 & 97.75 & 97.35 & -0.4 \\  
        Brighton and Hove & 48 & 23 & -25 & 97.79 & 97.78 & -0.01 \\  
        Darlington & 49 & 60 & 11 & 97.92 & 99.16 & 1.24 \\  
        Portsmouth & 50 & 69 & 19 & 97.99 & 99.51 & 1.52 \\  
        Wirral & 51 & 62 & 11 & 98 & 99.17 & 1.18 \\  
        Lancashire & 52 & 61 & 9 & 98.16 & 99.17 & 1.01 \\  
        Telford and Wrekin & 53 & 34 & -19 & 98.16 & 98.35 & 0.19 \\  
        Haringey & 54 & 57 & 3 & 98.18 & 99.07 & 0.89 \\  
        Westminster & 55 & 5 & -50 & 98.31 & 95.77 & -2.54 \\  
        North Lincolnshire & 56 & 50 & -6 & 98.36 & 98.82 & 0.45 \\  
        Bury & 57 & 77 & 20 & 98.37 & 99.85 & 1.48 \\  
        Isle of Wight & 58 & 67 & 9 & 98.37 & 99.35 & 0.98 \\  
        Sheffield & 59 & 85 & 26 & 98.38 & 100.02 & 1.64 \\  
        Northumberland & 60 & 65 & 5 & 98.46 & 99.32 & 0.86 \\  
        Cornwall & 61 & 32 & -29 & 98.51 & 98.32 & -0.19 \\  
        North Tyneside & 62 & 73 & 11 & 98.52 & 99.61 & 1.09 \\  
        Bournemouth, Christchurch and Poole& 63 & 46 & -17 & 98.55 & 98.73 & 0.18 \\  
        Tower Hamlets & 64 & 53 & -11 & 98.66 & 98.94 & 0.28 \\  
        Lincolnshire & 65 & 38 & -27 & 98.66 & 98.45 & -0.22 \\  
        Leeds & 66 & 86 & 20 & 98.69 & 100.05 & 1.35 \\  
        Coventry & 67 & 102 & 35 & 98.71 & 100.95 & 2.24 \\  
        Peterborough & 68 & 81 & 13 & 98.72 & 99.89 & 1.17 \\  
        East Sussex & 69 & 71 & 2 & 98.75 & 99.55 & 0.8 \\  
        Wigan & 70 & 83 & 13 & 98.76 & 99.96 & 1.2 \\  
        Slough & 71 & 42 & -29 & 98.79 & 98.61 & -0.18 \\  
        Lambeth & 72 & 30 & -42 & 98.8 & 98.23 & -0.57 \\  
        Kirklees & 73 & 93 & 20 & 98.83 & 100.5 & 1.67 \\  
        Norfolk & 74 & 33 & -41 & 98.88 & 98.35 & -0.53 \\  
        Thurrock & 75 & 78 & 3 & 98.96 & 99.85 & 0.9 \\  
        Southend-on-Sea & 76 & 70 & -6 & 98.98 & 99.53 & 0.54 \\  
        Luton & 77 & 89 & 12 & 99.05 & 100.22 & 1.17 \\  
        Calderdale & 78 & 79 & 1 & 99.25 & 99.87 & 0.61 \\  
        Swindon & 79 & 80 & 1 & 99.45 & 99.87 & 0.42 \\  
        Bedford & 80 & 82 & 2 & 99.57 & 99.89 & 0.33 \\  
        Camden & 81 & 35 & -46 & 99.64 & 98.37 & -1.27 \\  
        Lewisham & 82 & 90 & 8 & 99.67 & 100.25 & 0.58 \\  
        Southwark & 83 & 58 & -25 & 99.79 & 99.15 & -0.64 \\  
        Cumbria & 84 & 54 & -30 & 99.84 & 98.98 & -0.86 \\  
        Croydon & 85 & 91 & 6 & 99.9 & 100.47 & 0.57 \\  
        Newham & 86 & 105 & 19 & 99.95 & 101.08 & 1.13 \\  
        Brent & 87 & 111 & 24 & 99.96 & 101.36 & 1.4 \\  
        Herefordshire, County of & 88 & 40 & -48 & 100.03 & 98.51 & -1.52 \\  
        Waltham Forest & 89 & 99 & 10 & 100.07 & 100.82 & 0.74 \\  
        Hounslow & 90 & 87 & -3 & 100.11 & 100.06 & -0.05 \\  
        Hammersmith and Fulham & 91 & 72 & -19 & 100.29 & 99.55 & -0.74 \\  
        Hillingdon & 92 & 100 & 8 & 100.36 & 100.84 & 0.48 \\  
        Somerset & 93 & 74 & -19 & 100.49 & 99.69 & -0.8 \\  
        Northamptonshire & 94 & 94 & 0 & 100.55 & 100.62 & 0.07 \\  
        North Somerset & 95 & 88 & -7 & 100.6 & 100.19 & -0.41 \\  
        Enfield & 96 & 106 & 10 & 100.61 & 101.08 & 0.47 \\  
        Kent & 97 & 113 & 16 & 100.66 & 101.52 & 0.86 \\  
        Warrington & 98 & 108 & 10 & 100.7 & 101.28 & 0.58 \\  
        Derbyshire & 99 & 101 & 2 & 100.86 & 100.87 & 0.01 \\  
        Dorset & 100 & 68 & -32 & 101.09 & 99.41 & -1.68 \\  
        Staffordshire & 101 & 98 & -3 & 101.1 & 100.79 & -0.31 \\  
        Stockport & 102 & 114 & 12 & 101.14 & 101.64 & 0.5 \\  
        Suffolk & 103 & 92 & -11 & 101.14 & 100.47 & -0.67 \\  
        Milton Keynes & 104 & 116 & 12 & 101.14 & 101.76 & 0.62 \\  
        Essex & 105 & 109 & 4 & 101.43 & 101.33 & -0.1 \\  
        Ealing & 106 & 115 & 9 & 101.44 & 101.64 & 0.2 \\  
        Bexley & 107 & 126 & 19 & 101.52 & 102.5 & 0.98 \\  
        West Sussex & 108 & 97 & -11 & 101.56 & 100.78 & -0.77 \\  
        Worcestershire & 109 & 107 & -2 & 101.66 & 101.2 & -0.46 \\  
        Shropshire & 110 & 59 & -51 & 101.66 & 99.16 & -2.51 \\  
        East Riding of Yorkshire & 111 & 103 & -8 & 101.79 & 101.01 & -0.78 \\  
        Reading & 112 & 127 & 15 & 101.81 & 102.55 & 0.74 \\  
        Nottinghamshire & 113 & 120 & 7 & 101.82 & 102.02 & 0.2 \\  
        Gloucestershire & 114 & 104 & -10 & 101.82 & 101.02 & -0.8 \\  
        Sutton & 115 & 125 & 10 & 101.83 & 102.43 & 0.6 \\  
        Cheshire West and Chester & 116 & 117 & 1 & 101.85 & 101.76 & -0.09 \\  
        Solihull & 117 & 124 & 7 & 101.99 & 102.38 & 0.39 \\  
        Kensington and Chelsea & 118 & 76 & -42 & 102.2 & 99.83 & -2.37 \\  
        Cambridgeshire & 119 & 95 & -24 & 102.4 & 100.66 & -1.74 \\  
        Devon & 120 & 96 & -24 & 102.58 & 100.74 & -1.84 \\  
        Havering & 121 & 135 & 14 & 102.73 & 103.2 & 0.47 \\  
        York & 122 & 119 & -3 & 103 & 101.93 & -1.08 \\  
        Warwickshire & 123 & 121 & -2 & 103.04 & 102.15 & -0.9 \\  
        Redbridge & 124 & 133 & 9 & 103.09 & 103.17 & 0.08 \\  
        Wiltshire & 125 & 112 & -13 & 103.27 & 101.5 & -1.77 \\  
        Wandsworth & 126 & 122 & -4 & 103.3 & 102.19 & -1.11 \\  
        North Yorkshire & 127 & 110 & -17 & 103.44 & 101.35 & -2.09 \\  
        Bath and North East Somerset & 128 & 129 & 1 & 103.51 & 102.81 & -0.7 \\  
        Barnet & 129 & 136 & 7 & 103.53 & 103.21 & -0.32 \\  
        Merton & 130 & 137 & 7 & 103.59 & 103.29 & -0.3 \\  
        Central Bedfordshire & 131 & 123 & -8 & 103.64 & 102.32 & -1.32 \\  
        Leicestershire & 132 & 118 & -14 & 103.68 & 101.9 & -1.78 \\  
        Bromley & 133 & 139 & 6 & 104 & 103.71 & -0.29 \\  
        South Gloucestershire & 134 & 130 & -4 & 104.02 & 102.97 & -1.05 \\  
        Cheshire East & 135 & 132 & -3 & 104.12 & 103.12 & -1 \\  
        Trafford & 136 & 134 & -2 & 104.2 & 103.19 & -1.01 \\  
        Oxfordshire & 137 & 131 & -6 & 104.54 & 102.97 & -1.56 \\  
        Hertfordshire & 138 & 138 & 0 & 104.74 & 103.68 & -1.06 \\  
        Harrow & 139 & 144 & 5 & 104.88 & 104.46 & -0.42 \\  
        Hampshire & 140 & 142 & 2 & 105.2 & 104.28 & -0.92 \\  
        Surrey & 141 & 140 & -1 & 105.69 & 104.1 & -1.59 \\  
        Buckinghamshire & 142 & 141 & -1 & 105.86 & 104.26 & -1.61 \\  
        Kingston upon Thames & 143 & 146 & 3 & 105.87 & 105.34 & -0.53 \\  
        West Berkshire & 144 & 143 & -1 & 106.62 & 104.42 & -2.2 \\  
        Rutland & 145 & 128 & -17 & 106.63 & 102.77 & -3.87 \\  
        Bracknell Forest & 146 & 147 & 1 & 106.76 & 105.36 & -1.4 \\  
        Windsor and Maidenhead & 147 & 145 & -2 & 107.43 & 105.19 & -2.23 \\  
        Richmond upon Thames & 148 & 149 & 1 & 108.57 & 107.1 & -1.47 \\  
        Wokingham & 149 & 148 & -1 & 109.33 & 106.58 & -2.76 \\  \hline
        \caption{Ranking and Index comparison between ONS original version and the modified version.} \label{tab:long1} \\
    \end{longtable}

\begin{table}
  \centering
    \begin{threeparttable}[b]
  \caption{\label{tab:SM.Var}Variance explained by the first component by years.}
\begin{tabular}{llcccc}
\toprule 
Subdomains & Description & 2015 & 2016 & 2017 & 2018 \\ 
\midrule
L1.1 &  Physiological risk factors & 71.02 & 71.53 & 71.70 & 71.74 \\
  Li.2 & Behavioural risk factors  &58.18 & 55.23 & 52.26 & 58.24 \\ 
  Li.3 & Unemployment &  100.00 & 100.00 & 100.00 & 100.00 \\ 
  Li.4 & Working conditions & 54.88 & 52.00 & 51.73 & 50.04 \\ 
  Li.5 &Risk factors for children& 54.11 & 55.86 & 55.02 & 55.45 \\ 
  Li.6 & Children and young people’s education& 63.09 & 63.92 & 69.06 & 69.74 \\ 
  Li.7 &Protective measures& 80.25 & 80.74 & 83.17 & 86.65 \\ 
  Pe.1 & Mortality&  94.57 & 94.56 & 93.92 & 93.04 \\ 
  Pe.2 &Physical health conditions & 81.37 & 80.86 & 81.19 & 80.18 \\ 
  Pe.3 & Difficulties in daily life& 75.01 & 78.64 & 77.37 & 75.04 \\ 
  Pe.4 & Personal well-being &72.16 & 70.15 & 73.28 & 72.69 \\ 
  Pe.5 &Mental health & 68.54 & 66.07 & 65.97 & 68.08 \\ 
  Pl.1 & Access to green space& 70.67 & 70.67 & 70.67 & 70.67 \\ 
  Pl.2 & Local environment& 69.21 & 70.72 & 70.71 & 71.22 \\ 
  Pl.3 & Access to housing& 68.88 & 65.40 & 68.65 & 69.84 \\ 
  Pl.4 &Access to services& 93.53 & 94.31 & 94.34 & 94.36 \\ 
  Pl.5 & Crime & 100.00 & 100.00 & 100.00 & 100.00 \\ 
\bottomrule
\end{tabular}
\end{threeparttable}
\end{table}

 \begin{table}
 \centering
 \begin{threeparttable}
\caption{\label{tab:tabesti} Estimates for the  first  $S_i$ and total $S_{T_i}$ order sensitivity for the Health Index and the domains}
 \begin{tabular}{cccc}
  Domain & Step          & Sensitivity Index & Estimate 95\%CI \\\hline
Index  & Winsorization & First Order       & 0.01(0,0.02)     \\
     &               & Total Order       & 0.08(0.07,0.08)  \\
       & Normalization & First Order       & 0.01(0,0.02)     \\
       &               & Total Order       & 0.05(0.04,0.05)  \\
       & Aggregation   & First Order       & 0.86(0.81,0.91)  \\
       & Weights       & Total Order       & 0.98(0.95,1)     \\ \hline
 Lives  & Winsorization & First Order       & 0(-0.02,0.02)    \\
     &               & Total Order       & 0.16(0.16,0.17)  \\
       & Normalization & First Order       & 0.78(0.73,0.82)  \\
       &               & Total Order       & 0.93(0.91,0.95)  \\
      & Aggregation   & First Order       & 0.05(0.03,0.07)  \\
       & Weights       & Total Order       & 0.1(0.1,0.11)    \\\hline
People & Winsorization & First Order       & 0(0,0)           \\
       &               & Total Order       & 0(0,0)           \\
       & Normalization & First Order       & 0.75(0.72,0.78)  \\
       &               & Total Order       & 0.81(0.79,0.83)  \\
      & Aggregation   & First Order       & 0.19(0.18,0.21)  \\
       & Weights       & Total Order       & 0.22(0.21,0.23)  \\ \hline
 Places & Winsorization & First Order       & 0.27(0.24,0.31)  \\
       &               & Total Order       & 0.33(0.33,0.34)  \\
      & Normalization & First Order       & 0.68(0.64,0.72)  \\
      &               & Total Order       & 0.74(0.72,0.75)  \\
      & Aggregation   & First Order       & 0(0,0.01)        \\
 & Weights       & Total Order       & 0.01(0.01,0.01)   \\
\bottomrule
\end{tabular}
\end{threeparttable}
\end{table}

\clearpage

\clearpage
\section*{Additional Figures}

\begin{figure}[ht]
\centering
 \makebox{\includegraphics[scale=0.7]{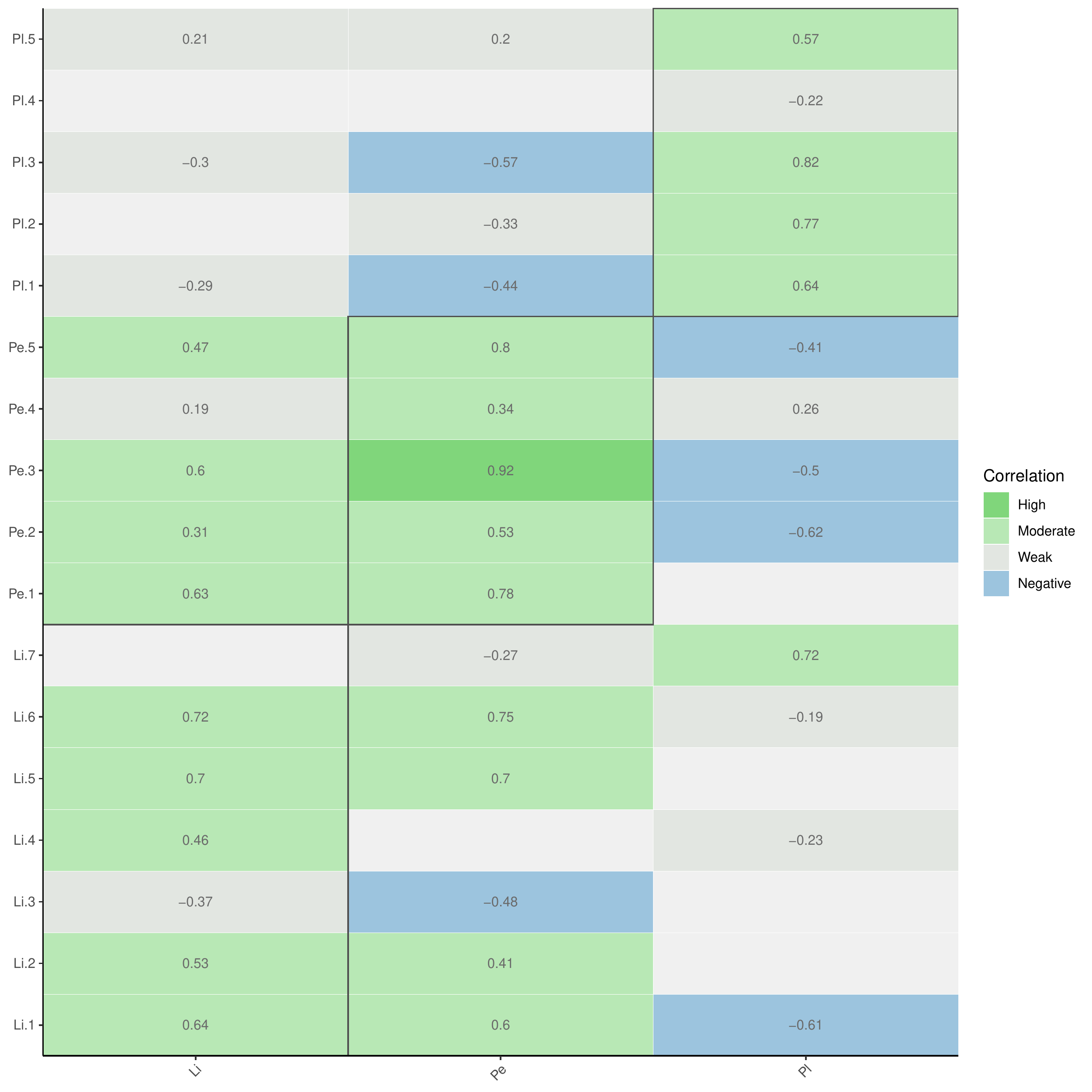}}
\caption{\label{fig:DomCorr1}Correlation between subdomains and domains Lives (LI), People(Pe), Places(Pl), grouped by subdomains.}
\end{figure}

\clearpage

\begin{figure}[ht]
 \centering
 \makebox{\includegraphics[scale=0.55]{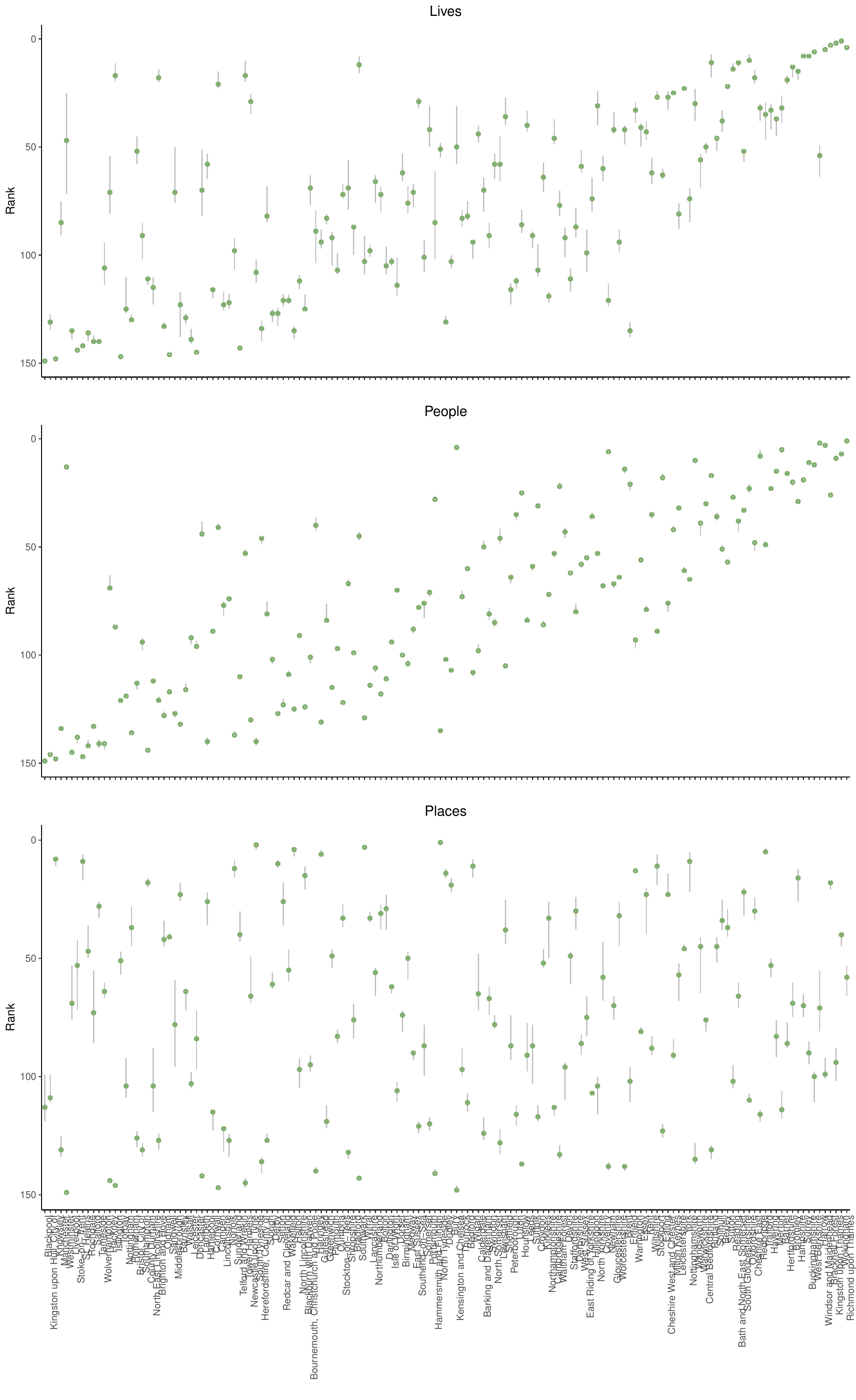}}
 \caption{\label{fig:figSAD}Results of UA showing the three domains, with the UTLA ordered as the median ranking for the overall index, with  the corresponding 5\textsuperscript{th} and 95\textsuperscript{th} percentiles (bounds). }
\end{figure}

\clearpage

\begin{figure}[ht]
\centering
\includegraphics[width=1\textwidth]{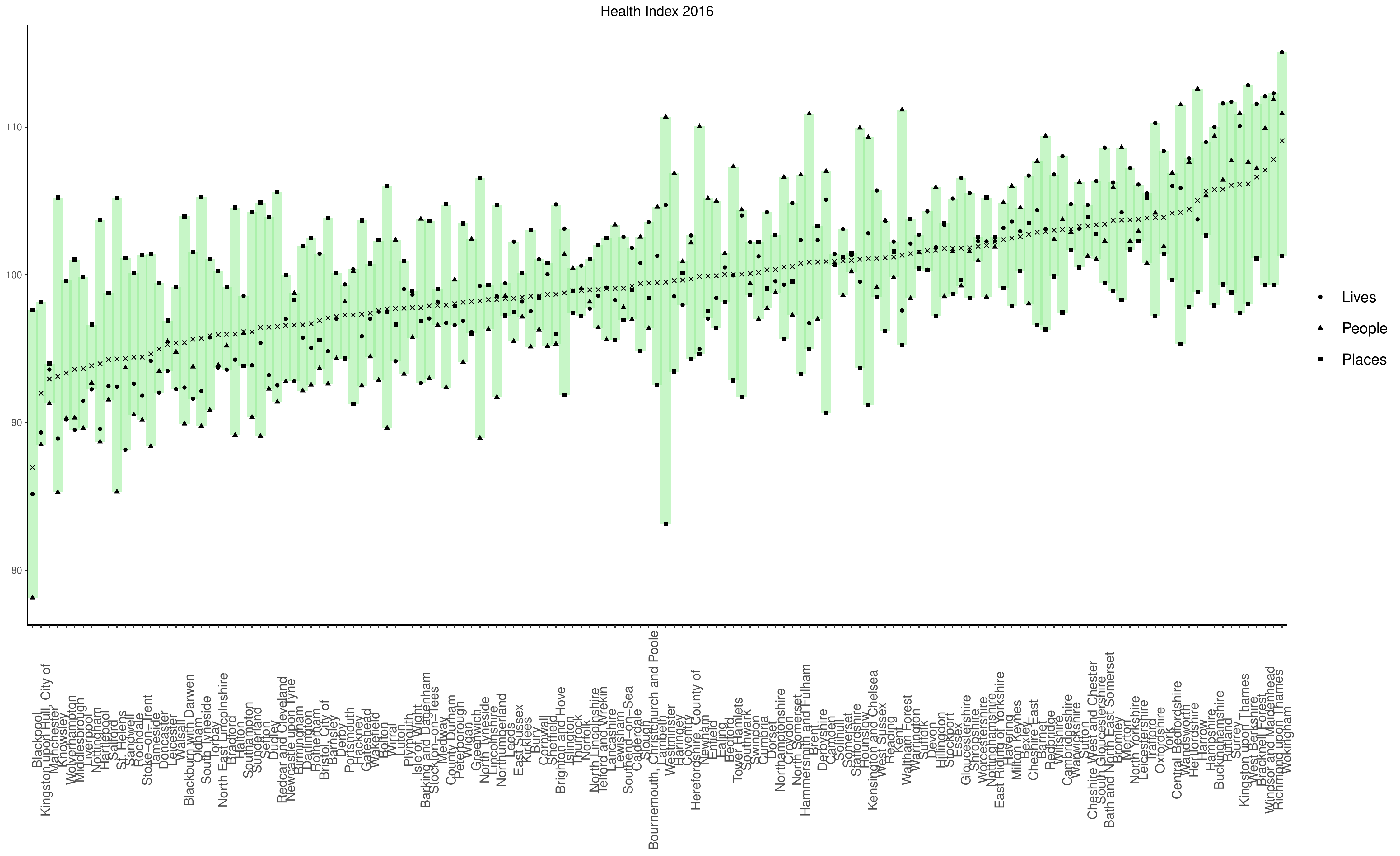}
\caption{\label{fig:figS1} UTLA Ranking ONS Health Index 2016 }
\end{figure}

\begin{figure}[ht]
\centering
\includegraphics[width=1\textwidth]{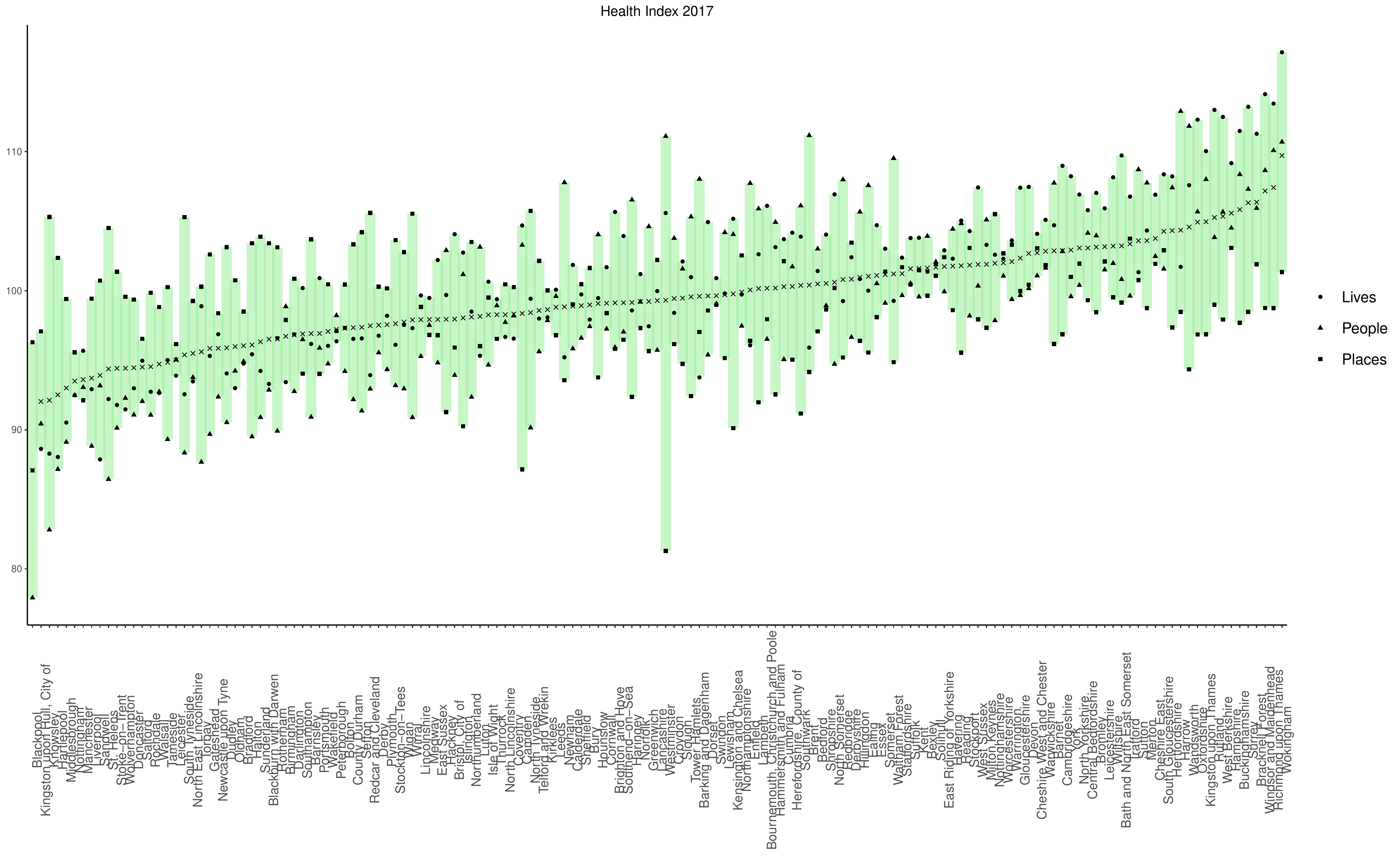}
\caption{\label{fig:figS2} UTLA Ranking ONS Health Index 2017 }
\end{figure}

\begin{figure}[ht]
\centering
\includegraphics[width=1\textwidth]{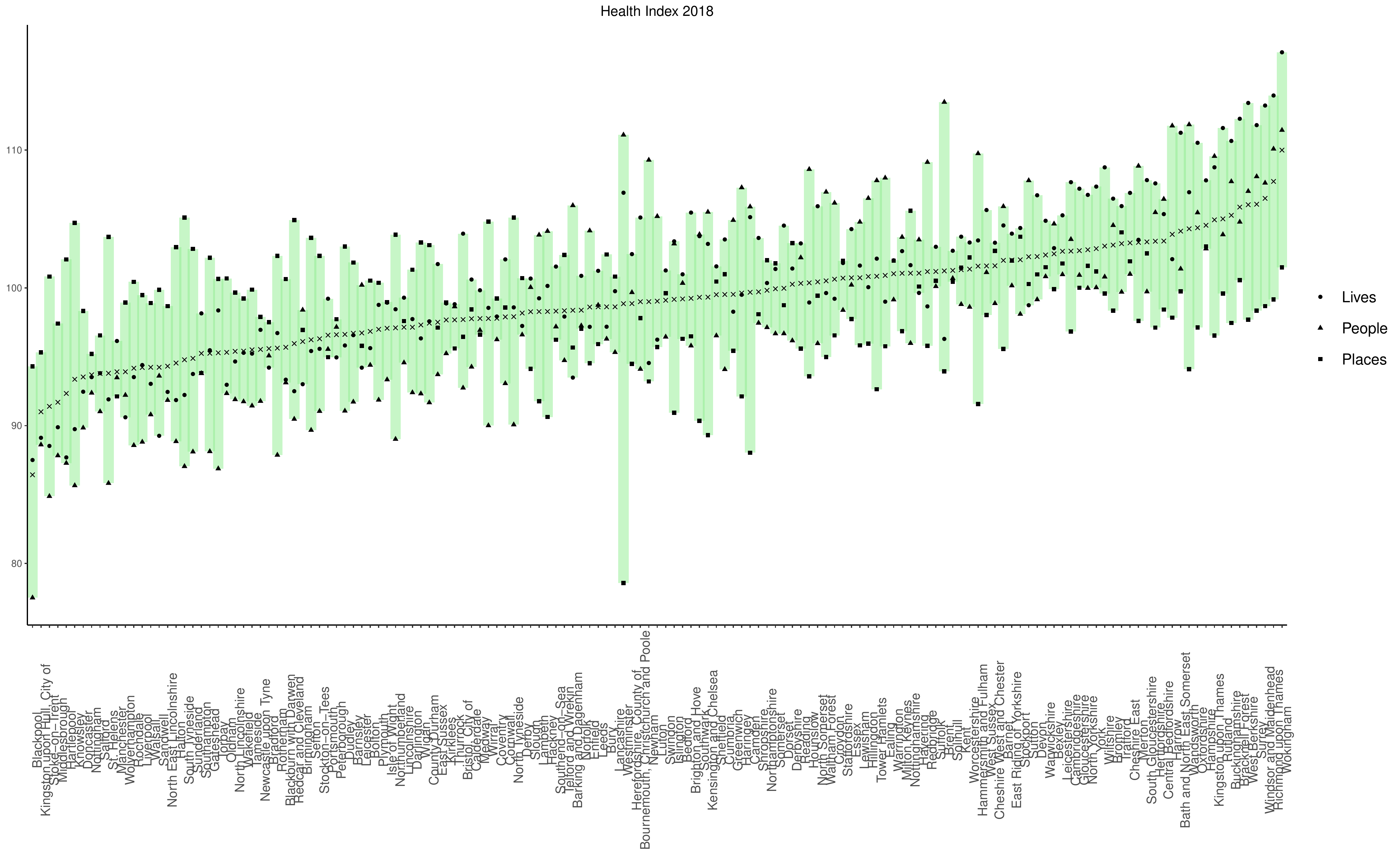}
\caption{\label{fig:figS3} UTLA Ranking ONS Health Index 2018}
\end{figure}

\newpage

\bibliographystyle{alpha}
\bibliography{compsiteb}